\documentclass[12pt]{article}
\usepackage[dvips]{graphicx}
\usepackage{amsmath}
\input{amssym.tex}
\newlength{\dinwidth}
\newlength{\dinmargin}
\begin{document}
\vspace{1 cm}
\newcommand{\address}{ } 
\newcommand{\gev}       {\mbox{${\rm Gev}$}}
\newcommand{\gevsq}     {\mbox{${\rm GeV}^2$}}
\newcommand{\qsd}       {\mbox{${Q^2}$}}
\newcommand{\PTM}       {P_T\hspace{-2.2ex}/\hspace{1.2ex}}
\newcommand{\x}         {\mbox{${\it x}$}}
\newcommand{\y}         {\mbox{${\it y}$}}
\newcommand{\ye}        {\mbox{${y_{e}}$}}
\newcommand{\smallqsd}  {\mbox{${q^2}$}}
\newcommand{\ra}        {\mbox{$ \rightarrow $}}
\newcommand{\ygen}      {\mbox{${y_{gen}}$}}
\newcommand{\yjb}       {\mbox{${y_{_{JB}}}$}}
\newcommand{\yda}       {\mbox{${y_{_{DA}}}$}}
\newcommand{\qda}       {\mbox{${Q^2_{_{DA}}}$}}
\newcommand{\qjb}       {\mbox{${Q^2_{_{JB}}}$}}
\newcommand{\ypt}       {\mbox{${y_{_{PT}}}$}}
\newcommand{\qpt}       {\mbox{${Q^2_{_{PT}}}$}}
\newcommand{\ypr}       {\mbox{${y_{(1)}}$}}
\newcommand{\yprpr}     {\mbox{${y_{(2)}}$}}
\newcommand{\ypps}      {\mbox{${y_{(2)}^2}$}}
\newcommand{\gammah}    {\mbox{$\gamma_{_{H}}$}}
\newcommand{\gammahc}   {\mbox{$\gamma_{_{PT}}$}}
\newcommand{\gap}       {\hspace{0.5cm}}
\newcommand{\gsim}      {\mbox{\raisebox{-0.4ex}{$\;\stackrel{>}{\scriptstyle 
\sim}\;$}}}
\newcommand{\lsim}      {\mbox{\raisebox{-0.4ex}{$\;\stackrel{<}{\scriptstyle 
\sim}\;$}}}
\newcommand{\ptrat}     {\mbox{$\frac{p_{Th}}{p_{Te}}$}}
\newcommand{\yjbrat}    {\mbox{$\frac{y_{_{JB}}}{y_{gen}}$}}
\newcommand{\ydarat}    {\mbox{$\frac{y_{_{DA}}}{y_{gen}}$}}
\newcommand{\yptrat}    {\mbox{$\frac{y_{_{PT}}}{y_{gen}}$}}
\newcommand{\yprrat}    {\mbox{$\frac{y_{(1)}}{y_{gen}}$}}
\newcommand{\yprprrat}  {\mbox{$\frac{y_{(2)}}{y_{gen}}$}}
\newcommand{\yerat}     {\mbox{$\frac{y_{e}}{y_{gen}}$}}
\newcommand{\yptye}     {\mbox{$\frac{y_{_{PT}}}{y_{e}}$}}
\newcommand{\qptrat}    {\mbox{$\frac{Q^2_{PT}}{Q^2_{gen}}$}}
\newcommand{\qdarat}    {\mbox{$\frac{Q^2_{DA}}{Q^2_{gen}}$}}
\newcommand{\qerat}     {\mbox{$\frac{Q^2_{e}}{Q^2_{gen}}$}}
\newcommand{\qprprrat}  {\mbox{$\frac{Q^2_{(2)}}{Q^2_{gen}}$}}
\def\3{\ss}
\def\ctr#1{{\it #1}\\\vspace{10pt}}
\renewcommand{\thefootnote}{\arabic{footnote}}
%
\newcommand{\lumi}      {47.7~pb$^{-1}$}
%
%
\newlength{\alskip}
\settowidth{\alskip}{$^{  45}$}
\setlength{\parindent}{0pt}
\setlength{\parskip}{5pt plus 2pt minus 1pt}
%
%
\title {\begin{flushright}{\large DESY--99--059}  \end{flushright}
\vspace*{2cm}\bf  
           Measurement of High-\boldmath${Q^2}$
   Charged-Current ${e^+p}$ Deep Inelastic Scattering Cross Sections 
                at HERA
        }
\author{ZEUS Collaboration}
\date{}
\maketitle\thispagestyle{empty}
\vspace{3 cm}
%
%
\begin{abstract}

The $e^+p$ charged-current deep inelastic scattering cross sections,
$d\sigma/dQ^2$ for $Q^2$ between 200 and 60000~GeV$^2$, and
$d\sigma/dx$ and $d\sigma/dy$ for $Q^2 > 200$~GeV$^2$, have been measured
with the ZEUS detector at HERA.
A data sample of \lumi, collected at a center-of-mass energy of 300~GeV,
has been used.  The cross section $d\sigma/dQ^2$ falls by a factor of
about 50000 as $Q^2$ increases from 280 to 30000~GeV$^2$.
The double differential cross section $d^2\sigma/dxdQ^2$ has also been
measured.
A comparison between the data and Standard Model (SM) predictions shows that
contributions from antiquarks ($\overline{u}$ and $\overline{c}$) and
quarks ($d$ and $s$) are both required by the data.
The predictions of the SM give a good description of the full body of the
data presented here.
A comparison of the charged-current cross section $d\sigma/dQ^2$ with
the recent ZEUS results for neutral-current scattering shows that
the weak and electromagnetic forces have similar strengths for $Q^2$
above $M^2_W, M^2_Z$.
A fit to the data for $d\sigma/dQ^2$ with the Fermi constant $G_F$ and $M_W$
as free parameters yields
    $G_F = \left( 
                     1.171      {\pm 0.034}       \,{\rm (stat.)} 
                               ^{+0.026}_{-0.032} \,{\rm (syst.)} 
                               ^{+0.016}_{-0.015} \,{\rm (PDF)} 
              \right)\times10^{-5}\,{\rm GeV^{-2}}$
and $M_W =        80.8         ^{+4.9}_{-4.5} \,{\rm (stat.)} 
                               ^{+5.0}_{-4.3} \,{\rm (syst.)} 
                               ^{+1.4}_{-1.3} \,{\rm (PDF)~GeV}$.
Results for $M_W$, where the propagator effect alone or the SM constraint
between $G_F$ and $M_W$ have been considered, are also presented.

\end{abstract}

\pagestyle{plain}
\thispagestyle{empty}
\clearpage
\pagenumbering{Roman}

\begin{center}                                                                                     
{                      \Large  The ZEUS Collaboration              }                               
\end{center}                                                                                       
  J.~Breitweg,                                                                                     
  S.~Chekanov,                                                                                     
  M.~Derrick,                                                                                      
  D.~Krakauer,                                                                                     
  S.~Magill,                                                                                       
  B.~Musgrave,                                                                                     
  A.~Pellegrino,                                                                                   
  J.~Repond,                                                                                       
  R.~Stanek,                                                                                       
  R.~Yoshida\\                                                                                     
 {\it Argonne National Laboratory, Argonne, IL, USA}~$^{p}$
% ...
\newpage                                                                                           
% ...
$^{\    2}$ also at IROE Florence, Italy \\                                                        
% ...
$^{  36}$ partially supported by the Foundation for German-Russian Collaboration                   
DFG-RFBR \\ \hspace*{3.5mm} (grant no. 436 RUS 113/248/3 and no. 436 RUS 113/248/2)\\              
% ...
                                                           %
                                                           %
\newpage   
                                                           %
                                                           %
\begin{tabular}[h]{rp{14cm}}                                                                       
$^{a}$ &  supported by the Natural Sciences and Engineering Research                               
          Council of Canada (NSERC)  \\                                                            
$^{b}$ &  supported by the FCAR of Qu\'ebec, Canada  \\                                            
$^{c}$ &  supported by the German Federal Ministry for Education and                               
          Science, Research and Technology (BMBF), under contract                                  
          numbers 057BN19P, 057FR19P, 057HH19P, 057HH29P, 057SI75I \\                              
$^{d}$ &  supported by the MINERVA Gesellschaft f\"ur Forschung GmbH, the                          
German Israeli Foundation, and by the Israel Ministry of Science \\                                
$^{e}$ &  supported by the German-Israeli Foundation, the Israel Science                           
          Foundation, the U.S.-Israel Binational Science Foundation, and by                        
          the Israel Ministry of Science \\                                                        
$^{f}$ &  supported by the Italian National Institute for Nuclear Physics                          
          (INFN) \\                                                                                
$^{g}$ &  supported by the Japanese Ministry of Education, Science and                             
          Culture (the Monbusho) and its grants for Scientific Research \\                         
$^{h}$ &  supported by the Korean Ministry of Education and Korea Science                          
          and Engineering Foundation  \\                                                           
$^{i}$ &  supported by the Netherlands Foundation for Research on                                  
          Matter (FOM) \\                                                                          
$^{j}$ &  supported by the Polish State Committee for Scientific Research,                         
          grant No. 115/E-343/SPUB/P03/154/98, 2P03B03216, 2P03B04616,                             
          2P03B10412, 2P03B05315, 2P03B03517, and by the German Federal                            
          Ministry of Education and Science, Research and Technology (BMBF) \\                     
$^{k}$ &  supported by the Polish State Committee for Scientific                                   
          Research (grant No. 2P03B08614 and 2P03B06116) \\                                        
$^{l}$ &  partially supported by the German Federal Ministry for                                   
          Education and Science, Research and Technology (BMBF)  \\                                
$^{m}$ &  supported by the Fund for Fundamental Research of Russian Ministry                       
          for Science and Edu\-cation and by the German Federal Ministry for                       
          Education and Science, Research and Technology (BMBF) \\                                 
$^{n}$ &  supported by the Spanish Ministry of Education                                           
          and Science through funds provided by CICYT \\                                           
$^{o}$ &  supported by the Particle Physics and                                                    
          Astronomy Research Council \\                                                            
$^{p}$ &  supported by the US Department of Energy \\                                              
$^{q}$ &  supported by the US National Science Foundation \\                                       
$^{r}$ &  partially supported by the British Council,                                              
          ARC Project 0867.00 \\                                                                   
$^{s}$ &  partially supported by the British Council,                                              
          Collaborative Research Project, TOK/880/11/15                                            
\end{tabular}                                                                                      
                                                           %
%
%
\newpage
\pagestyle{plain}
\pagenumbering{arabic}                   
\setcounter{page}{1}
\setlength{\parindent}{3.4ex}
\setlength{\parskip}{0pt}
%
%
\section{\bf Introduction}
\label{s:intro}

\begin{figure}
  \begin{center}
    $ \begin{array}{c c }
      \begin{minipage}{6cm}
        \includegraphics*[width=6cm]{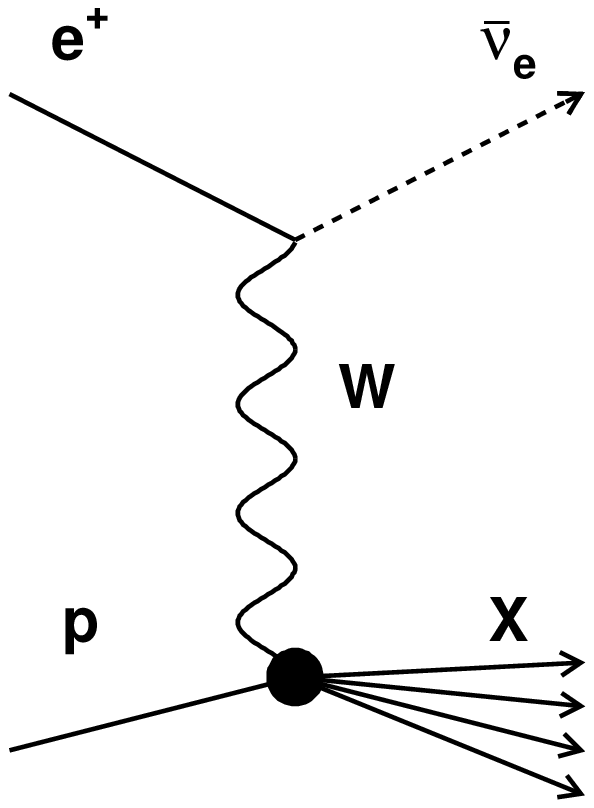}
      \end{minipage}
      &
      \begin{minipage}{99mm}
        \includegraphics*[bb=69 387 500 659,width=99mm]{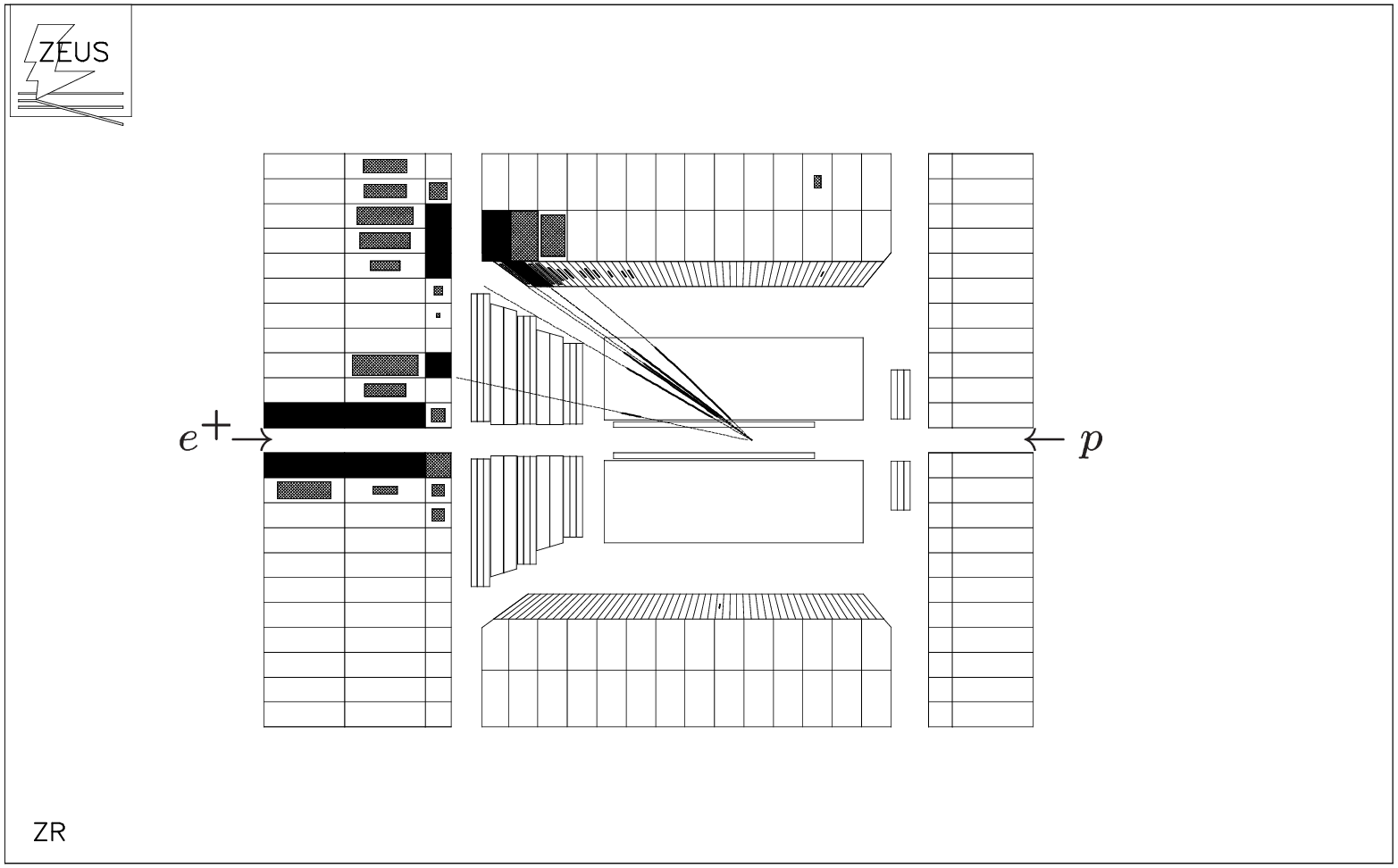}
      \end{minipage}
      \\
      (a) & (b) 
    \end{array}$
  \end{center}
  \caption{(a) A schematic diagram of charged-current positron-proton
    scattering.  (b) A view of a charged-current candidate event in the ZEUS
    detector, projected in the plane parallel to the beam.
    The filled boxes indicate energy deposits in the calorimeter.
    The transverse momentum imbalance can be clearly seen in the calorimeter
    and also from the tracks of charged particles measured in the central
    tracking detector.}
  \label{f:feyn}
\end{figure}

Deep inelastic scattering (DIS) of leptons on nucleons
is the key source of information for the development of our understanding
of the structure of the nucleon.
In the Standard Model (SM),
charged-current (CC) DIS is mediated by the exchange of the
$W$ boson (see Fig.~\ref{f:feyn}(a)).
In contrast to neutral-current (NC) interactions, where all quark and
antiquark flavors contribute, only down-type quarks (antiquarks)
and up-type antiquarks (quarks)
participate at leading order in $e^+p$ ($e^-p$) CC DIS reactions.
Therefore, CC DIS provides a powerful tool for the
flavor-specific investigation of parton momentum distributions.
Even though CC events are kinematically less constrained than NC events
due to the unobserved final-state neutrino, they can be identified
with little background at HERA.

First measurements of the CC DIS cross section at HERA,
reported previously by the H1~\cite{H1CC,H1NCCC} and
ZEUS~\cite{ZEUSNCCC,ZEUSCC} collaborations, extended the coverage of the
kinematic range
compared to that of the fixed-target neutrino-nucleus scattering
experiments~\cite{NuScat} by about two orders of magnitude in the
four-momentum transfer squared ($-Q^2$).
These analyses were based on $e^-p$ and $e^+p$ data samples of approximately
1~pb$^{-1}$ and 3~pb$^{-1}$, respectively.
The cross section at high $Q^2$ demonstrated, for the first time, the presence
of a space-like propagator with a finite mass, consistent with that of the
$W$ boson.

This paper presents results from ZEUS on the CC $e^+p$ DIS
differential cross sections $d\sigma/dQ^2$, $d\sigma/dx$, $d\sigma/dy$ and
$d^2\sigma/dxdQ^2$ for $Q^2>200$~GeV$^2$, and comparisons to SM predictions.  
The measurements are based on \lumi~of data collected
with the ZEUS detector from
1994\,--\,1997 during which HERA collided 27.5~GeV positrons with 820~GeV
protons, yielding a center-of-mass energy $\sqrt{s}~=~300$~GeV.
The 16-fold increase in the luminosity compared to the previous
measurements allows the double differential cross
section $d^2\sigma/dxdQ^2$ to be determined
in this high-$Q^2$ regime for the first time.
A recent publication presented NC cross sections from the same
data sample~\cite{ZNCpaper}.
These data, together with those presented here, permit a precise comparison
of CC and NC cross sections up to $Q^2$ values of about
$2{\cdot}10^4$~GeV$^2$.
%
%
\section{\bf Standard Model prediction}
\label{s:KineSM}

   The electroweak Born cross section for the reaction 
\begin{equation}
e^+ p \rightarrow \bar{\nu}_e X
\end{equation}
can be written as
\begin{equation}
  {{d^2\sigma^{\rm CC}_{\rm Born}(e^+p)} \over {dx \, dQ^2}} =
  {G^2_F \over 4 \pi x } \Biggl( {M^2_W \over M^2_W + Q^2}\Biggr)^2
  \Biggl[
Y_+ { F}^{\rm CC}_2(x, Q^2) - Y_- x{ F}^{\rm CC}_3(x, Q^2) - y^2{ F}^{\rm CC}_{ L}(x, Q^2)
  \Biggr],
  \label{e:Born}
\end{equation}
where $G_F$ is the Fermi constant, $M_W$ is the mass of the $W$ boson,
$x$ is the Bjorken scaling variable, $y=Q^2/xs$ and
$Y_\pm = 1 \pm \left( 1 - y \right)^2$.
The center-of-mass energy of the positron-proton collision is given by
$\sqrt{s} = 2\sqrt{E_eE_p}$,
where $E_e$ and $E_p$ are the positron and proton beam energies, respectively.
The structure functions ${ F}^{\rm CC}_2$ and 
$x{ F}^{\rm CC}_3$ , in leading-order (LO) QCD, measure
sums and differences of quark and antiquark parton momentum
distributions \cite{Devenish}.  For longitudinally unpolarized
beams,
\begin{equation}
{ F}^{\rm CC}_2 = x[d(x, Q^2) + s(x, Q^2) + \bar{u}(x, Q^2) + \bar{c}(x, Q^2)],
  \label{e:F2}
\end{equation}
\begin{equation}
x{ F}^{\rm CC}_3 = x[d(x, Q^2) + s(x, Q^2) - \bar{u}(x, Q^2) - \bar{c}(x, Q^2)],
  \label{e:F3}
\end{equation}
where $d(x,Q^2)$ is, for example,
the parton distribution function (PDF) which gives the number density
of a down quark with momentum fraction $x$ in the proton.
Since the top quark mass is large and the off-diagonal elements of
the CKM matrix are small, the contribution from the third generation
quarks to the structure functions may be safely ignored~\cite{uli}.
The chirality of the CC interaction is reflected by the factors $Y_\pm$
multiplying the structure functions.
The longitudinal structure function, $F^{\rm CC}_{L}$,
is zero at leading order but is finite at next-to-leading-order (NLO) QCD.
It gives a negligible contribution to the cross section except at
$y$ values close to 1, where it can be as large as 10\%.

   The electroweak radiative corrections to (\ref{e:Born}) receive
contributions from initial state photon radiation,
fermion and boson loops, and the exchange of multiple intermediate
vector bosons.
The effects of these radiative corrections are taken
into account to leading order~\cite{HCL},
so that the quoted cross sections in this paper
are corrected to the electroweak Born level.
Equation (\ref{e:Born}) is evaluated
with $G_F = 1.16639 \times 10^{-5}~\rm{GeV}^{-2}$
and $M_W = 80.41~\rm{GeV}$ \cite{PDG}.
The uncertainties in the electroweak parameters
have a negligible effect both on the
calculated cross sections and on the radiative corrections.

   Thus, the main uncertainty in the SM cross-section prediction
comes from the PDF uncertainties, which are
discussed in detail in~\cite{ZNCpaper,botje} and are
taken into account in the CC cross-section calculation.
The resulting uncertainty in $d\sigma/dQ^2$, for example, ranges
from 4\% at $Q^2=200$~GeV$^2$ to 10\% at $Q^2=10000$~GeV$^2$, and increases
further at higher $Q^2$.
The large uncertainty at high $Q^2$ is due to the $d$-quark density
which is poorly constrained at high $x$ by the experimental data.
%
%
\section{\bf The ZEUS experiment}
\label{s:detector}

ZEUS~\cite{detector} is a multipurpose magnetic detector designed 
to measure $ep$ interactions at HERA.  The primary components used 
for this analysis are the compensating uranium-scintillator 
calorimeter (CAL), the central tracking detector (CTD), and the luminosity 
detector.

The ZEUS coordinate system is right-handed with the $Z$ axis pointing in the
direction of the proton beam (forward) and the $X$ axis pointing horizontally
toward the center of HERA.
The polar angle $\theta$ is zero in the $Z$ direction.

Tracking information is provided by the CTD~\cite{b:CTD} operating in 
a 1.43 T solenoidal magnetic field.  The interaction vertex
is measured with a typical resolution along (transverse to)
the beam direction of 0.4~(0.1)~cm.
The CTD is used to reconstruct the momenta of 
tracks in the polar angle region $15^\circ < \theta < 164^\circ$.  
The transverse momentum ($p_t$)
resolution for full-length tracks can be parameterized 
as $\sigma(p_t)/p_t=0.0058\ p_t \oplus 0.0065 \oplus 0.0014/p_t$,
with $p_t$ in GeV.

The CAL~\cite{b:CAL} covers $99.7\%$ of the total solid angle. 
It is divided into three parts with a corresponding division in 
$\theta$ as viewed from the nominal interaction point: 
forward (FCAL, $2.6^\circ < \theta < 36.7^\circ$), barrel
(BCAL, $36.7^\circ < \theta < 129.1^\circ$) and rear (RCAL,
$129.1^\circ < \theta < 176.2^\circ$). Each section is subdivided into
towers which subtend solid angles between 0.006 and 0.04 steradian.
Each tower is longitudinally segmented into an electromagnetic (EMC) and 
one (RCAL) or two (FCAL, BCAL) hadronic sections (HAC). The electromagnetic
section of each tower is further subdivided transversely into two
(RCAL) or four (BCAL, FCAL) cells. Under test beam conditions the
calorimeter resolutions were $\sigma/E = 18\%/\sqrt{E(\rm GeV)}$ for
electrons and $\sigma/E = 35\%/\sqrt{E(\rm GeV)}$ for hadrons. The
calorimeter has a time resolution of better than 1~ns for energy deposits
above 4.5~GeV.
The position of the interaction vertex along the beam direction
can also be reconstructed from the measured arrival time of
energy deposits in FCAL~\cite{1992F2}.
The resolution is about 9~cm for events with FCAL energy above 25~GeV and
improves to about 7~cm for FCAL energy above 100~GeV.

An instrumented-iron backing calorimeter~\cite{b:BAC}
(BAC) measures energy leakage from the CAL.
The muon chambers in the forward~\cite{detector},
barrel and rear~\cite{b:BRMU} regions are used in this analysis
to detect background events induced by cosmic-ray or beam-halo muons.

The luminosity is measured using the Bethe-Heitler reaction
$ep \rightarrow ep\gamma$~\cite{b:LUMI}. The resulting small angle
energetic photons are measured by the luminosity monitor, a
lead-scintillator calorimeter placed in the HERA tunnel 107 m from the
interaction point in the positron beam direction.
%
%
\section{Monte Carlo simulation}
\label{s:Simu}

Monte Carlo simulations (MC) are used to determine the efficiency for 
selecting events, to determine the accuracy of kinematic 
reconstruction, to estimate the background rate and to extrapolate
measured cross sections to the full kinematic phase space.
A sufficient number of events is generated to ensure that errors from MC 
statistics can be neglected.
The MC samples are normalized to the total
integrated luminosity of the data.

The ZEUS detector response is simulated with a program based on 
{\sc geant}~\cite{geant}.  The generated events are passed through 
the simulated detector, subjected to the same trigger requirements as 
the data, and processed by the same reconstruction programs.

The underlying distribution of the $Z$-coordinate of the event vertex
is determined using a minimum-bias sample of low-$Q^2$
neutral-current DIS events as discussed in detail in \cite{ZNCpaper}.

CC DIS events including radiative effects are simulated 
using the {\sc heracles} 4.5.2~\cite{HCL} program with the 
{\sc django}6 2.4~\cite{DJANGO} interface to the QCD programs.  
In {\sc heracles}, corrections for initial-state radiation, vertex and 
propagator corrections, and two-boson exchange are included.
The QCD cascade and the hadronic final state are simulated
using the color-dipole model of {\sc ariadne} 4.08~\cite{Ariadne}
and, as a systematic check, the {\sc meps} model of
{\sc lepto} 6.5~\cite{Lepto}.
Both programs use the Lund string model of {\sc jetset} 7.4~\cite{JETSET}
for the hadronization.
A set of NC events generated with {\sc django} is used to estimate
the NC contamination in the CC sample.

Photoproduction background is estimated using events 
simulated with {\sc herwig}~\cite{Herwig}.
The background from $W$ production is estimated
using the {\sc epvec}~\cite{EPVEC} generator, and the background from
Bethe-Heitler production of charged-lepton pairs
is generated with the {\sc lpair}~\cite{LPAIR} program.
%
%
\section{Reconstruction of kinematic variables}
\label{s:Rec}
 
The principal signature of CC DIS events at HERA
is the presence of a large missing transverse momentum, $\PTM$.
This is illustrated in Fig.~\ref{f:feyn}(b), where an event from the final
CC DIS sample is shown.
The struck quark gives rise to one or more jets of hadrons.
The energetic final-state neutrino escapes detection, leaving a large
imbalance in the transverse momentum observed in the detector.
$\PTM$ is calculated as
\begin{equation}
\PTM^2  =  P_x^2 + P_y^2 = 
  \left( \sum\limits_{i} E_i \sin \theta_i \cos \phi_i \right)^2
+ \left( \sum\limits_{i} E_i \sin \theta_i \sin \phi_i \right)^2,
  \label{eq:pt}
\end{equation}
where the sum runs over all calorimeter energy deposits $E_i$ 
(uncorrected in the trigger, but corrected in the
offline analysis as discussed below), and
$\theta_i$ and $\phi_i$ are their polar and azimuthal angles
as viewed from the interaction vertex.
The hadronic polar angle, $\gamma_h$, is defined by
\begin{equation}
\cos\gamma_h = \frac{\PTM^2 - \delta^2}{\PTM^2 + \delta^2},
\label{eq:gammah}
\end{equation}
where
\begin{equation}
\delta = \sum\limits_{i} ( E_i - E_i \cos \theta_{i} ) 
= \sum\limits_{i} (E-p_z)_{i}.
\label{eq:delta}
\end{equation}
In the na\"{\i}ve Quark Parton Model,
$\gamma_h$ gives the angle of the struck quark.
Another variable used in the selection is the total transverse energy,
$E_T$, given by
\begin{equation}
E_T    = \sum\limits_{i} E_i \sin \theta_i. \label{eq:et}
\end{equation}

The kinematic variables are reconstructed using the Jacquet-Blondel
method~\cite{JB}.  The estimators of $y$, $Q^2$ and $x$ are:
\begin{equation}
y_{JB} = \delta/(2E_e); {\hskip 1cm}
Q^2_{JB} = \PTM^2/(1-y_{JB}); {\hskip 1cm}
x_{JB} = Q^2_{JB}/(sy_{JB}).
\end{equation}

For the offline determination of $\PTM, \delta$ and $E_T$,
methods developed and tested for the NC cross section
determination~\cite{ZNCpaper} are used.
The calorimeter cells with energy deposits
are grouped into units called clusters.
For each cluster, corrections depending on the cluster energy and angle
are made for hadronic energy loss in inactive material in front of
the calorimeter.
The correction algorithm, which is based on MC, has been verified
using the highly constrained NC events measured in the ZEUS detector.
Energetic hadron jets in the FCAL direction may produce particles
backscattered into the BCAL or RCAL (albedo).
Also, particles may be redirected by the material between
the interaction point and the calorimeter.
Such effects, which create biases in the measurement of $\gamma_h$, are
suppressed by removing low energy clusters
at polar angles much larger than the calculated value of $\gamma_h$.
%
%
\section{Event selection}
\label{s:EvSel}
 
CC DIS candidates are selected by requiring large $\PTM$
and a reconstructed event vertex consistent with an $ep$ interaction.
The main sources of background affecting the CC event selection are processes
like NC DIS and high-$E_T$ photoproduction, where the finite resolution or
energy escaping detection in the CAL cause $\PTM$.
Events not originating from $ep$ collisions such as beam-gas interactions,
beam-halo muons or cosmic rays can also cause substantial apparent
imbalance in the transverse momentum and constitute
other sources of background.
The selection criteria described below are imposed
to separate CC events from the background.

The events are classified first according to $\gamma_0$, the value of
$\gamma_h$ measured with respect to the nominal interaction point.
If $\gamma_0$ is sufficiently large, i.e. in the central region,
tracks in the CTD are used to reconstruct the event vertex, which strongly
suppresses non-$ep$ backgrounds.
The selection procedure designed to select these events is described
in Sect.~\ref{ss:StanEvSel}.
On the other hand, if $\gamma_0$ is small, i.e. in the forward region,
the hadronic final state of such CC events
is often outside the acceptance of the CTD, and thus calorimeter timing
is used for the vertex reconstruction.
The algorithm designed specifically to select such events, which
tend to be at high $x$ values, is described in Sect.~\ref{ss:LowGEvSel}.
The kinematic quantities are finally recalculated using the
$Z$-coordinate of the event vertex ($Z_{\rm VTX}$) determined
from either CTD tracks or calorimeter timing, depending on $\gamma_0$.

\subsection{Trigger selection}
\label{ss:Trigger}

ZEUS has a three-level trigger system~\cite{detector}.
At the first trigger level, events are selected using criteria based on the
energy, transverse energy and missing transverse momentum
determined by the calorimeter~\cite{CFLT}.
Generally, events are triggered with a lower threshold of these values
in coincidence with at least one CTD track, while a higher threshold is
necessary for events with no CTD track.
The latter events have a hadronic final state boosted forward outside
the CTD acceptance.
Typical threshold values are 5~GeV (8~GeV) in missing transverse momentum,
or 11.5~GeV (30~GeV) in transverse energy, for events with (without)
CTD tracks.

At the second level, timing information from the calorimeter
is used to reject background events inconsistent with the bunch-crossing time.
Also, the missing transverse momentum is available with better resolution
than at the first level, so that a tighter
cut of 6~GeV (9~GeV without CTD track) can be made.

At the third level, track reconstruction and vertex finding are performed
and are used to reject candidate events with a vertex that is inconsistent
with an $ep$ interaction.
The thresholds on the trigger quantities are lower than the cut variables
used in the offline analysis.

\subsection{Offline selection based on a CTD vertex}
\label{ss:StanEvSel}

Events with $\gamma_{0}>23^\circ$ are required to contain a vertex
reconstructed from CTD tracks and to satisfy the following criteria:
\begin{itemize}
\item {$\PTM > 12$ GeV and $\PTM ' > 10$ GeV\\}
  $\PTM '$ is the missing transverse momentum calculated excluding the FCAL
  towers closest to the beam hole.  The $\PTM '$ cut strongly suppresses
  beam-gas background events while maintaining
  high efficiency for CC events.
\item {$| Z_{\rm VTX} | < 50$~cm\\}
  A vertex reconstructed by the CTD is required to be within the range
  consistent with the $ep$ interaction region.
\item {Tracking requirement\\}
  At least one track associated with the event vertex must have
  transverse momentum in excess of 0.2~GeV and 
  a polar angle in the range $15^\circ$ to $164^\circ$. 
\item {Rejection of photoproduction\\}
  Photoproduction events tend to have azimuthally symmetric hadronic
  energy flow.  At high $E_T$, a relatively small imbalance due to resolution
  effects or escaping particles can lead to non-negligible $\PTM$.
  These events are rejected by the following cuts:
  $\PTM/E_T > 0.4  $ is required for events with $20 < \PTM < 30$ GeV;
  $\PTM/E_T > 0.55  $ is required for events with $\PTM < 20$ GeV.
  No $\PTM/E_T$ requirement is imposed on events with $\PTM>30$~GeV.
  In addition, the difference between the
  direction of the $(P_x,P_y)$ vector calculated using CTD tracks and
  that obtained using the calorimeter is required to be less than 1~radian if
  $\PTM < 20$~GeV and less than 2~radians if $\PTM > 20$~GeV.
\item {Rejection of NC DIS\\}
  NC DIS events in which the positron or jet energy is poorly measured
  can have a large $\PTM$.  To identify such events, a positron-finding
  algorithm which selects isolated electromagnetic
  clusters~\cite{Sinistra} is used.
  Candidate positron clusters within the CTD acceptance are required to 
  have an energy above 4~GeV and
  a matching track with momentum larger than 25\% of the cluster energy.
  Clusters with $\theta > 164^\circ$ are required to have a transverse
  momentum exceeding 2~GeV.
  Events with a candidate positron satisfying the above criteria
  and $\delta > 30$~GeV are rejected; for contained NC events,
  $\delta$ peaks at $2E_e = 55$~GeV.
  This cut is applied only for events with $\PTM < 30$\,GeV.
\item {Rejection of non-$ep$ background\\}
  Beam-gas events typically give calorimeter arrival times which are
  inconsistent with the bunch-crossing time.  Such events are rejected.
  A muon-finding algorithm based on calorimeter energy deposits or
  muon-chamber signals is used to reject events produced by
  cosmic-ray or beam-halo muons.
\end{itemize}

\subsection{Offline selection without CTD vertex}
\label{ss:LowGEvSel}

Events with $\gamma_0<23^\circ$ are not required to have an event vertex
reconstructed from CTD tracks.
They must satisfy the following criteria:
\begin{itemize}
\item {$\PTM > 14$ GeV and $\PTM ' > 12$ GeV\\}
  Relaxing the requirements on tracking and the CTD vertex
  results in an increase of non-$ep$ background.
  To compensate for this, the requirements
  on the missing transverse momentum are tightened.
\item {$| Z_{\rm VTX} | < 50$~cm\\}
  $Z_{\rm VTX}$ is reconstructed from the
  measured arrival time of energy deposits in FCAL.
  The relation between the timing measurement
  and $Z_{\rm VTX}$ was determined using
  a large data sample of NC DIS events, in which a reliable $Z_{\rm VTX}$
  estimate can be obtained from the positron track even if the hadronic
  system is boosted in the very forward direction.
\item {Rejection of photoproduction\\}
  $ \PTM/E_T > 0.6 $ is required for events with $\PTM < 30$ GeV.
  This cut also suppresses beam-gas interactions.
\item {Rejection of non-$ep$ background\\}
  The same timing and muon-rejection cuts are used as described
  in Sect.~\ref{ss:StanEvSel}.
  A class of background events which are especially
  troublesome in this selection
  branch arises from beam-halo muons interacting inside the FCAL.
  To reduce this background,
  topological cuts on the transverse and longitudinal
  shower shape are imposed; these reject events where the energy
  deposits are much more strongly collimated than for
  typical hadronic jets.
  Another characteristic of muons traversing the detector
  parallel to the beam line
  is a coincidence of energy deposits in the RCAL and FCAL at similar
  $(X, Y)$ positions.
  If such a coincidence is found, the event is rejected.
\end{itemize}
NC DIS is negligible in this selection branch.

\subsection{Final event sample}
\label{ss:FidBin}

In order to restrict the sample to regions where
the resolution in the kinematic variables is acceptable and
the background is small, further requirements $Q^2_{JB}>200$~GeV$^2$
and $y_{JB}<0.9$ are imposed.
The cross sections presented below are
corrected to the full $y$ range using the SM $y$-dependence described
by (\ref{e:Born}).

The combined selection efficiency of the above cuts
for most of the $x$ and $y$ region ($0.1< x$, $0.1 < y < 0.9$)
is typically 90\%.
At low $x$ or high $y$, the efficiency decreases due to the $\PTM$
requirement.  At low $y$, the hadronic system is close
to the beam pipe and the $\PTM '$ requirement affects the efficiency.
The overall selection efficiency for CC events with $Q^2 > 200$~GeV$^2$
is 70\%.

\begin{figure}
  \begin{center}
    \includegraphics*[width=\textwidth]{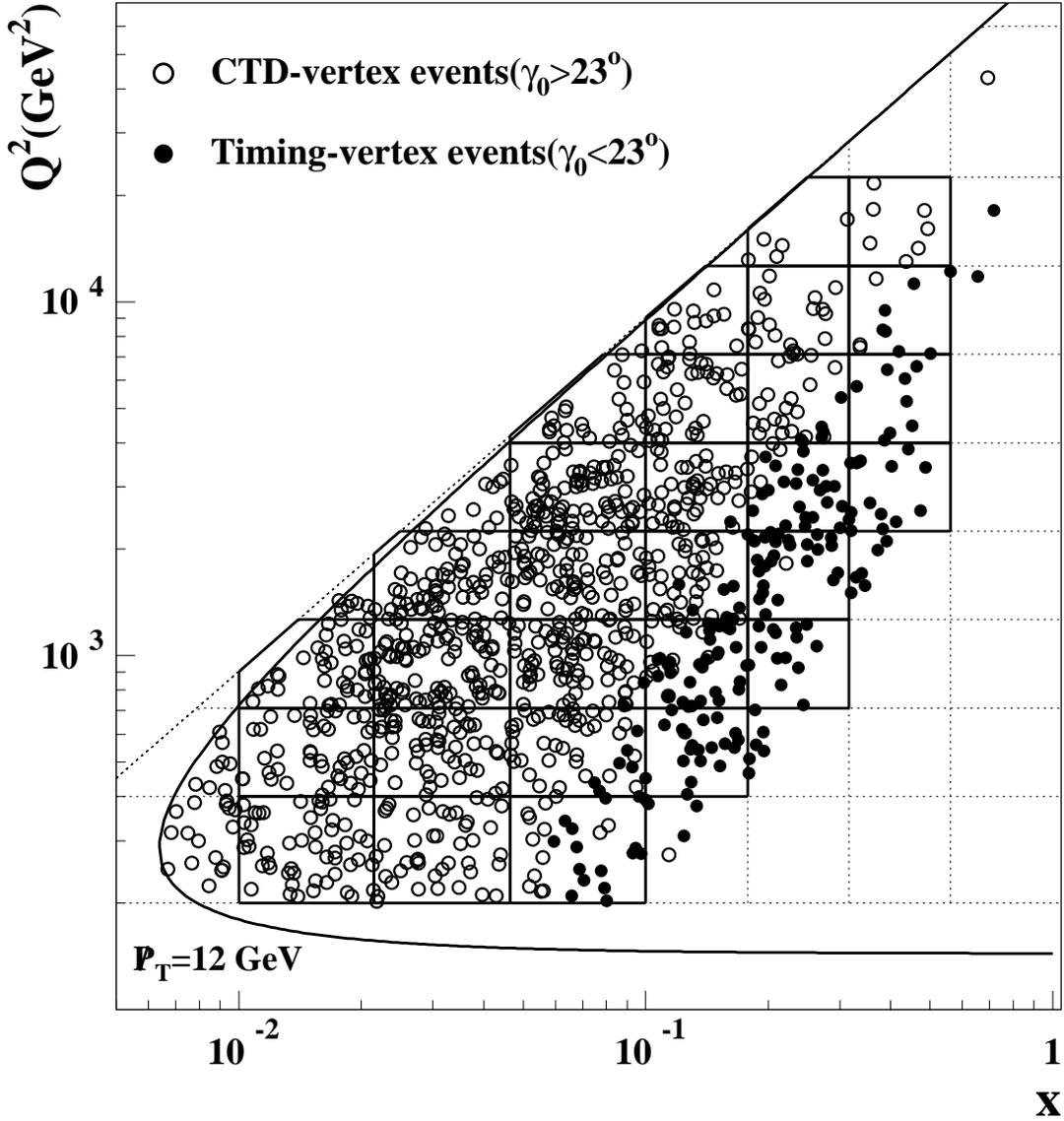}
  \end{center}
  \caption{Distribution of the selected CC candidates in the $(x,Q^2)$
    plane.
    Open (full) circles show the events selected with (without)
    tracking vertex.
    The curve shows the $\PTM$ cut of 12~GeV.
    The bin boundaries are shown by the dotted lines, delimited by 
    the diagonal dotted line of the kinematic limit, $y=1$.
    The bins used in the double differential cross section measurement
    are marked with solid lines.
}
  \label{f:scat}
\end{figure}

The final sample consists of 1086 events.
All events have been scanned visually, and no remaining cosmic or
halo-muon background events have been found.
The distribution of $Q^2$ versus $x$ for the accepted events
is shown in Fig.~\ref{f:scat}.
For $x>0.2$, the sample is dominated by events for which the
interaction vertex is determined by the calorimeter timing (full circles). 
Figure~\ref{f:scat} also demonstrates that the acceptance is zero
for events at very low x in the low $Q^2$ region due to the $\PTM$ cut.
There is also zero acceptance for events at
very low $y$ (low $Q^2$ and high $x$), where a large part of the
hadronic system escapes in the forward beam pipe direction.
The MC is used to correct for the acceptance loss of such events
in determining the cross sections.

\begin{figure}
  \begin{center}
    \includegraphics*[width=.75\textwidth]{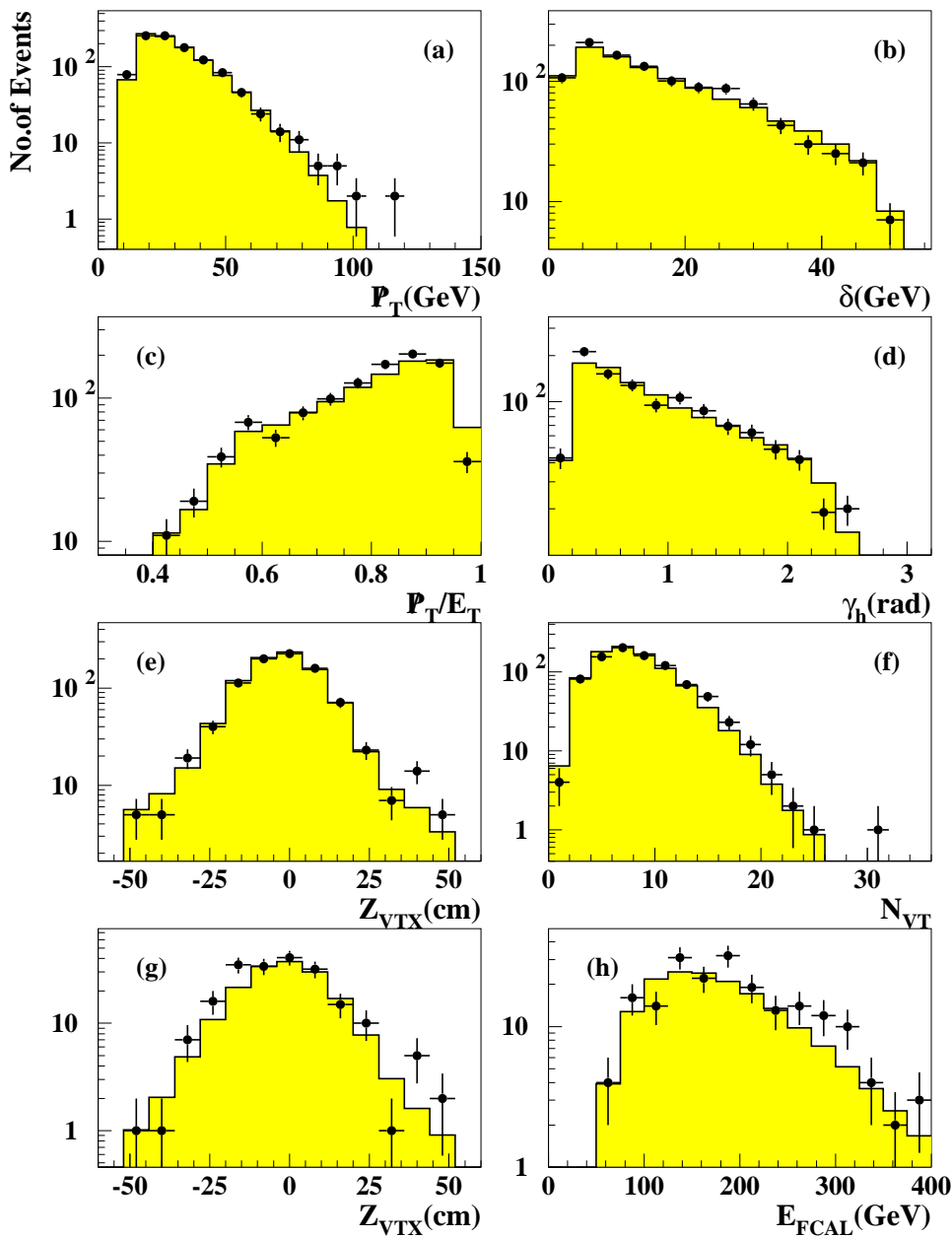}
  \end{center}
  \caption{
    Comparison of the final CC data sample (solid points) with the
    expectations of the MC (histograms), normalized to the
    luminosity of the data.
    The distributions of (a) the missing transverse momentum,
    $\PTM$, (b) the variable $\delta$, (c) $\PTM/E_T$,
    the ratio of missing transverse momentum to total transverse energy
    and (d) the variable $\gamma_h$, are shown.
    In (e) and (f), the distributions of the $Z$ position
    of the event vertex and the number of tracks assigned to the
    primary vertex, N$_{\rm VT}$, are shown for selected events with
    CTD vertex (see Sect.~\ref{ss:StanEvSel}).
    In (g) and (h), the distribution of $Z_{\rm VTX}$ and 
    $E_{\rm FCAL}$ are shown for events passing the selection
    with the timing vertex (see Sect.~\ref{ss:LowGEvSel}).
    }
  \label{f:MCvData}
\end{figure}

Figures~\ref{f:MCvData}(a)-(d) show the distributions
of the variables $\PTM$, $\delta$, $\PTM/E_T$ and $\gamma_h$
in the final event sample, compared with the corresponding MC predictions,
which include the contributions from CC DIS and the small contributions
from the background sources described below.
The CC MC distribution is based on the CTEQ4D~\cite{CTEQ4} PDF set (see
discussion in Sect.~\ref{sss:Single}).
The contamination from events not due to $ep$ collisions,
such as beam-gas interactions, is negligible.
In general, good agreement is observed, except for some excess
at low $\gamma_h$.
Furthermore, the peak of the $\PTM/E_T$ distribution is shifted to
slightly lower values as compared to MC.
For the events selected with a CTD vertex (Sect.~\ref{ss:StanEvSel}),
shown by the open circles in Fig.~\ref{f:scat},
the distributions of $Z_{\rm VTX}$ and the number of
tracks assigned to the primary vertex are provided in
Fig.~\ref{f:MCvData}(e) and (f).
The good agreement between data and MC shows that as far as the acceptance
calculation is concerned, the CC final state is well modeled by the MC.
For the events selected with a timing vertex (Sect.~\ref{ss:LowGEvSel}),
the distributions of $Z_{\rm VTX}$ and the total energy in FCAL, 
$E_{\rm FCAL}$, are shown in (g) and (h).
About 18\% of data events fall in this category, while MC predicts 16\%.
This small excess is directly related to the excess in the low
$\gamma_h$ (i.e. high $x$) region seen in (d).
As seen in (h), all events in this sample have large FCAL energies
to ensure a good resolution for $Z_{\rm VTX}$ from the timing.

The relative resolution in $Q^2$ is approximately 20\% over the
entire range of $Q^2$.
The relative resolution in $x$ improves from $\sim 20$\%
in the interval $0.01<x<0.0215$ (see Sect.~\ref{s:Xsect})
to $\sim 8\%$ at high $x$, and that in $y$ is approximately 8\% over the 
entire range, except for $y<0.1$ where it increases to $\sim 11$\%.
Here the resolutions are obtained by comparing the reconstructed quantities 
with the true values in MC, and the RMS value of the distribution is quoted.

The fraction of background events in the final sample is typically
below 1\% at high $Q^2$ and increases as $Q^2$ decreases,
exceeding 10\% in the interval
200~GeV$^2 < Q^2 <$ 400~GeV$^2$ (see Sect.~\ref{s:Xsect}),
as estimated from MC.
Photoproduction and Bethe-Heitler dilepton production in the dimuon channel
are the dominant sources of background at low $Q^2$,
whereas the production and decay of time-like $W$ bosons
is the remaining background at high $Q^2$.
The contamination from NC events is negligible.
%
%
\section{Cross-section determination}
\label{s:Xsect}

The single and double differential cross sections are determined using
bin-by-bin unfolding.  The measured cross section
in a particular bin, $\sigma_{\rm meas}$, is determined from
\begin{equation}
  \sigma_{\rm meas} = \frac{N_{\rm obs} - N_{\rm bg}}{\cal{A} \cal{L}},
  \label{e:Rad}
\end{equation}
where $N_{\rm obs}$ is the number of observed events in the bin, 
$N_{\rm bg}$ is the estimated number of background events, $\cal{A}$ is the
acceptance and $\cal{L}$ is the integrated luminosity.  The acceptance, 
defined from the MC as the number of events reconstructed within the bin
divided by the number of events generated in that bin, takes both
the selection efficiency and the event migration due to resolution
into account.

The measured cross section includes the radiative effects discussed in
Sect.~\ref{s:KineSM}.
The correction factor to provide the Born-level cross section is defined as
\begin{equation}
  \mathcal{C}_{\rm rad} = \frac{\sigma^{\rm SM}_{\rm Born}}
                               {\sigma^{\rm SM}_{\rm rad}}.
  \label{e:CRad}
\end{equation}
The numerator is obtained by numerically integrating (\ref{e:Born}) over
the bin.
The value of $\sigma^{\rm SM}_{\rm rad}$, the cross section in the bin
including radiation, is calculated using
{\sc heracles} 4.6.2~\cite{HCL462}.
The measured Born-level cross section is then given by
\begin{equation}
  \sigma_{\rm Born} = \sigma_{\rm meas} \mathcal{C}_{\rm rad}.
  \label{e:BornInt}
\end{equation}
Finally, the quoted differential cross section, for example $d\sigma / dQ^2$,
is calculated as
\begin{equation}
  \frac{d\sigma}{dQ^2} = \frac{\sigma_{\rm Born}}{\sigma^{\rm SM}_{\rm Born}}
                         \frac{d\sigma^{\rm SM}_{\rm Born}}{dQ^2} .
  \label{e:BornDiff}
\end{equation}
A similar procedure is used for $d\sigma/dx, d\sigma/dy$ and
$d^2\sigma/dxdQ^2$.
In this manner, the acceptance factor $\cal{A}$ is used to correct the effect
of all the selection cuts (Sect.~\ref{s:EvSel}) and the cross sections are
extrapolated to the full kinematic range.
In particular the MC is used to extrapolate beyond
the $y$-region restricted by the $y_{JB}<0.9$ cut.
The differential cross sections $d\sigma/dx$ and $d\sigma/dy$ are quoted
in the region $Q^2 > 200$~GeV$^2$.

For the $d\sigma/dQ^2$ measurement, nine bins are used between
$Q^2 = 200$ and 60000~GeV$^2$.
The bins have equal width in ${\rm log_{10}}Q^2$
between 400 and 22494~GeV$^2$ (four bins per decade),
while the lowest and highest $Q^2$ bins have somewhat larger width.
For $d\sigma/dx$, bins of equal
width in ${\rm log_{10}}x$ are used, three between $x = 0.01$ and $x = 0.1$,
and four between $x = 0.1$ and $x = 1.0$.
For the $d\sigma/dy$ measurement, the $y$ region is divided equally into
two bins between $y = 0.0$ and
$y = 0.2$, and five bins between $y = 0.2$ and $y = 0.9$.
All bins are defined such that their sizes significantly exceed the
resolutions of the respective variables (see Sect.~\ref{ss:FidBin}).
The values at which $d\sigma/dQ^2$ and $d\sigma/dx$ are quoted,
$Q^2_c$ and $x_c$, are chosen to be near the logarithmic center of each bin,
except in the highest $Q^2$ and $x$ bins, where they are chosen
lower than the logarithmic center, reflecting the very steeply
falling cross sections.
The cross section $d\sigma/dy$ is quoted at
the center $y_c$ of each bin.

The statistical errors are calculated using the square root of the number
of measured events, $N$, for $N>100$ and otherwise from $68\%$ Poisson
confidence intervals around $N$.

The values of $Q^2_c$, $x_c$ and $y_c$, 
the number of observed events, $N_{\rm obs}$, the estimated number of 
background events, $N_{\rm bg}$, the acceptance, ${\cal{A}}$, 
and the radiative correction factor, $\mathcal{C}_{\rm rad}$, are given
in Tables~\ref{tab:dsdq}\,--\,\ref{tab:ds2dxdq2_2}.
%
%
\section{Systematic uncertainties}
\label{s:SystErr}

The major sources of systematic uncertainties
for the quoted cross sections are described below.
The individual uncertainties are added in quadrature separately
for the positive and negative deviations from the
nominal cross section values to obtain the total systematic uncertainty.
The uncertainty on the luminosity of the combined 1994\,--1997 sample
is 1.6\% and is not included in the total systematic uncertainty.
\begin{itemize}
\item {Uncertainty of the calorimeter energy scale\\}  
  The uncertainty in the hadronic energy scale of the calorimeter is
  determined by the methods described in \cite{ZNCpaper} to be
  2\% for FCAL and BCAL, and 3\% for RCAL.
  Varying the energy scale of the calorimeter separately by
  this amount in the detector simulation induces shifts of
  the Jacquet-Blondel estimators for the kinematic variables.
  The resulting systematic errors in the
  measured cross sections are typically less than
  $10\%$, but increase to $\sim 30\%$ in the highest $Q^2$ bin and
  $\sim 25\%$ in the highest $x$ bin.  

  A 2\% fraction of the accepted events have a measurable energy leakage
  from the CAL into the BAC.
  The average energy leakage for these events is 4~GeV.
  This effect on the cross-section measurement is negligible.
\item {Variation of selection thresholds\\}
  The threshold values of the selection cuts
  are varied independently in MC by typically 10\%.
  The largest effect is observed when varying the $\PTM$ threshold;
  this changes the cross section by
  8\% in the lowest $Q^2$ bin and 4\% in the highest $y$ bin.
  Varying the $\PTM '$ threshold produces a 3\% change
  in the lowest $y$ bin.
\item {Uncertainty in the parton-shower scheme\\}
  To test the sensitivity of the results to the details of the simulation of
  higher-order QCD effects in the hadronic final state,
  the {\sc lepto} {\sc meps} model is used instead of the 
  {\sc ariadne} model for calculating the acceptance.
  The largest effects are observed in the bins of
  lowest $Q^2$ (8\%), highest $Q^2$ (6\%),
  lowest $x$ (5\%) and highest $y$ (5\%).
\item {Background subtraction\\}
  The uncertainty of the photoproduction background is estimated
  by fitting a linear combination of the $\PTM/E_T$ distributions
  of the signal and the photoproduction MC samples to the corresponding
  distribution in the data, allowing the normalizations of the direct
  and resolved photoproduction components to vary.
  No cut on $\PTM/E_T$ is applied for this check.
  A 40\% uncertainty in the photoproduction background is found,
  leading to a sizable systematic error in the lowest $Q^2$
  bins (4\% in $d\sigma/dQ^2$ and maximum 7\% in $d^2\sigma/dxdQ^2$).
  At $Q^2>400$\,GeV$^2$, the effect is less than 1\%.
\item {Trigger efficiency\\}
  The simulation of the efficiency of the first-level trigger as a function
  of $\PTM$ is examined by using data events triggered independently.
  This independent trigger is efficient for CC events and
  is based on the calorimeter energy sums.
  The difference between the efficiencies calculated from the
  data and from MC has a negligible effect on the measured cross section.
\item {Choice of parton distribution functions\\} 
  The CC MC events are generated using the CTEQ4D PDFs.
  To examine the influence of variations of the PDFs on the
  cross-section measurement, the PDFs of MRSA~\cite{MRSA}
  and GRV94~\cite{GRV} are also considered.
  In addition, a modification of
  the $d$-quark to $u$-quark density ratio
  according to the prescription
  $(d/u) + \delta(d/u) = (d/u) + 0.1x(x+1)$
  (see~\cite{bodek-yang} and discussion in Sect.~\ref{ss:Results})
  has been tested.
  This increases the predicted $e^+p$ CC cross section at high $x$.

  The MC events are re-weighted using these
  alternative PDFs and new acceptance correction factors
  are computed.
  The change in the measured cross section is typically $< 1\%$ 
  and at most 2.5\% over the entire kinematic range of interest.

  The radiative correction factors also depend on the choice of the
  PDFs.  Using the program {\sc hector}~\cite{HECTOR},
  the difference in the radiative correction factors by changing the PDFs
  is found to be typically less than 0.1\% and thus is neglected.
  
  The calculation of the differential cross section
  in a given bin according to (\ref{e:BornDiff})
  uses the ratio of the differential to the
  integrated SM cross sections.
  The ratio is sensitive to the shape of the PDFs within this bin.
  The largest effects are observed in the highest $Q^2$ bin (3.5\%)
  and highest $x$ bin (3\%).
\item {The effect of $F_L$\\} 
  The {\sc django} program neglects the $F_L$ contribution
  when generating CC events.
  The corresponding 
  effect on the acceptance correction factors is evaluated by
  reweighting MC events with the ratio of the cross sections
  with and without $F_L$.  The largest effect is observed in the
  highest $y$ bin where it amounts to 1.5\%.
\item {Uncertainty in the radiative correction\\} 
  The uncertainties in the radiative correction as determined with
  {\sc heracles} are estimated to be smaller than 3\% in the kinematic
  region considered~\cite{HS_Durham} and are not included in the total
  systematic uncertainty.
\end{itemize}
%
%
\section{Cross-section results}
\label{ss:Results}

\subsection{Single differential cross sections}
\label{sss:Single}

\begin{figure}
  \begin{center}
    \includegraphics[height=19cm]{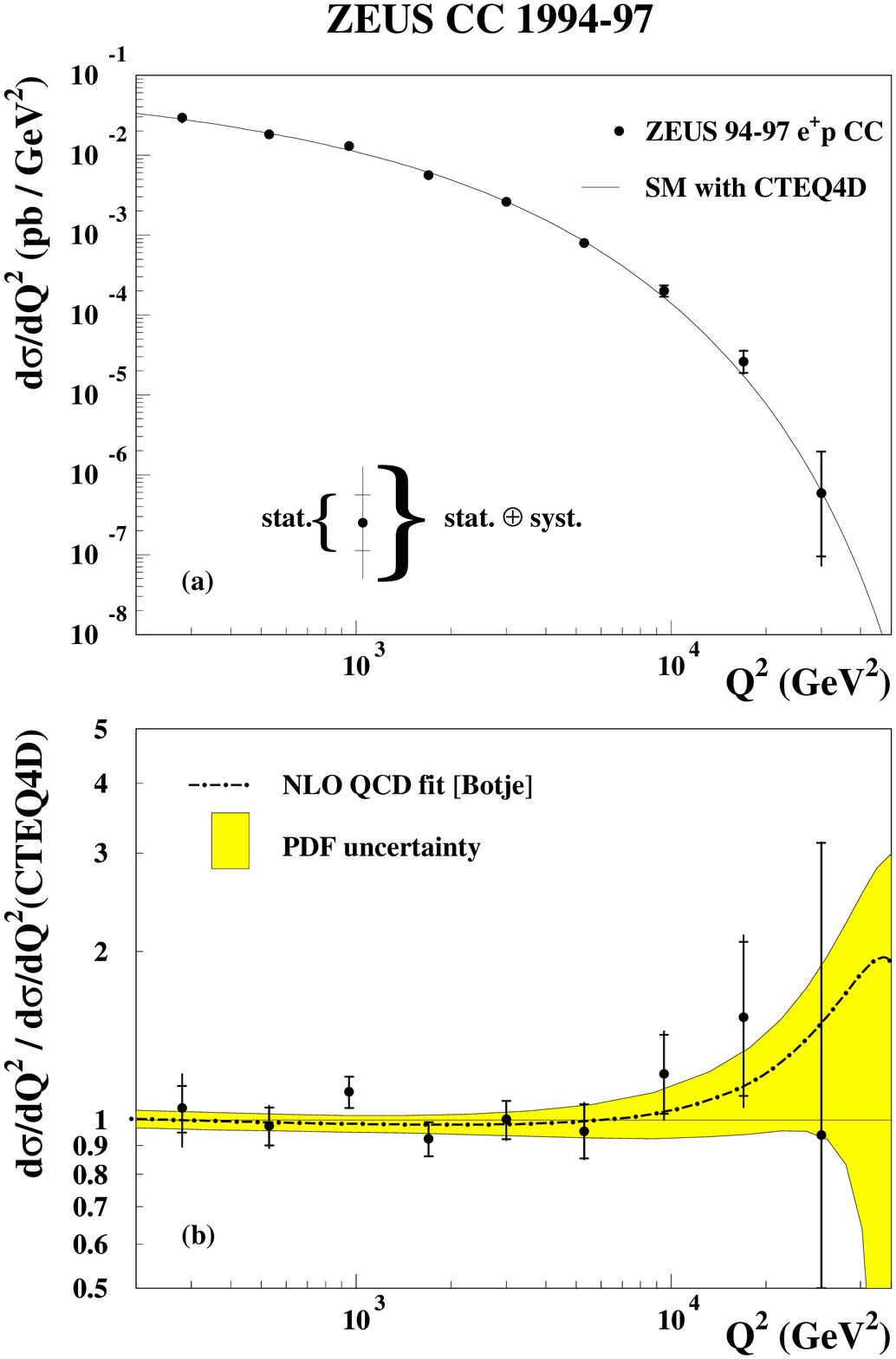}
  \end{center}
  \vskip -5mm
  \caption{
    (a) The $e^+p$ CC DIS Born cross section $d\sigma/dQ^2$ for data
    (solid points) and the Standard Model (SM) expectation evaluated using
    the CTEQ4D PDFs.
    (b) The ratio of the measured cross section $d\sigma/dQ^2$ to the
    SM expectation evaluated using the CTEQ4D PDFs.
    The statistical errors are indicated by the inner error bars (delimited by
    horizontal lines), while the full
    error bars show the total error obtained by adding the statistical and 
    systematic contributions in quadrature.
    Also shown by a dot-dashed line is the result of the NLO QCD fit
    together with the associated PDF uncertainties (shaded band).
    }
  \label{f:dsdq}
\end{figure}

\begin{figure}
  \begin{center}
    \includegraphics[width=.8\textwidth]{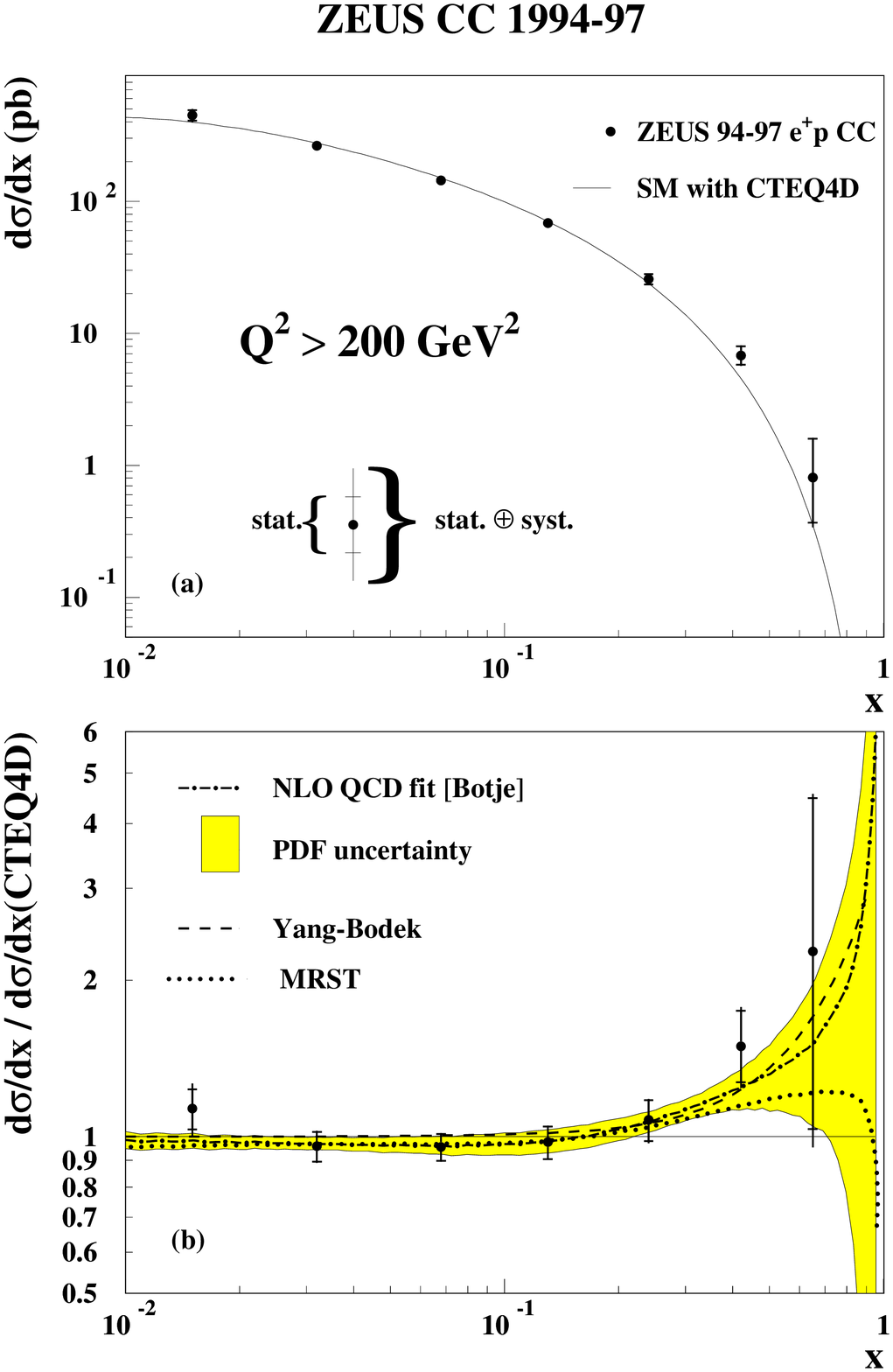}
  \end{center}
  \caption{ 
    (a) The $e^+p$ CC DIS Born cross section $d\sigma/dx$ for data
    (solid points) and the Standard Model (SM) expectation evaluated using
    the CTEQ4D PDFs.
    (b) The ratio of the measured cross section $d\sigma/dx$ to the
    SM expectation evaluated using the CTEQ4D PDFs.
    The statistical errors are indicated by the inner error bars (delimited by
    horizontal lines), while the full
    error bars show the total error obtained by adding the statistical and 
    systematic contributions in quadrature.
    Also shown by a dot-dashed line is the result of the NLO QCD fit
    together with the associated PDF uncertainties (shaded band).
    The dashed line represents the result of modifying the $d/u$ ratio with
    $\delta(d/u)=0.1x(x+1)$.  The dotted line shows the prediction from MRST
    PDFs.
    }
  \label{f:dsdx}
\end{figure}

\begin{figure}
  \begin{center}
    \includegraphics[width=.8\textwidth]{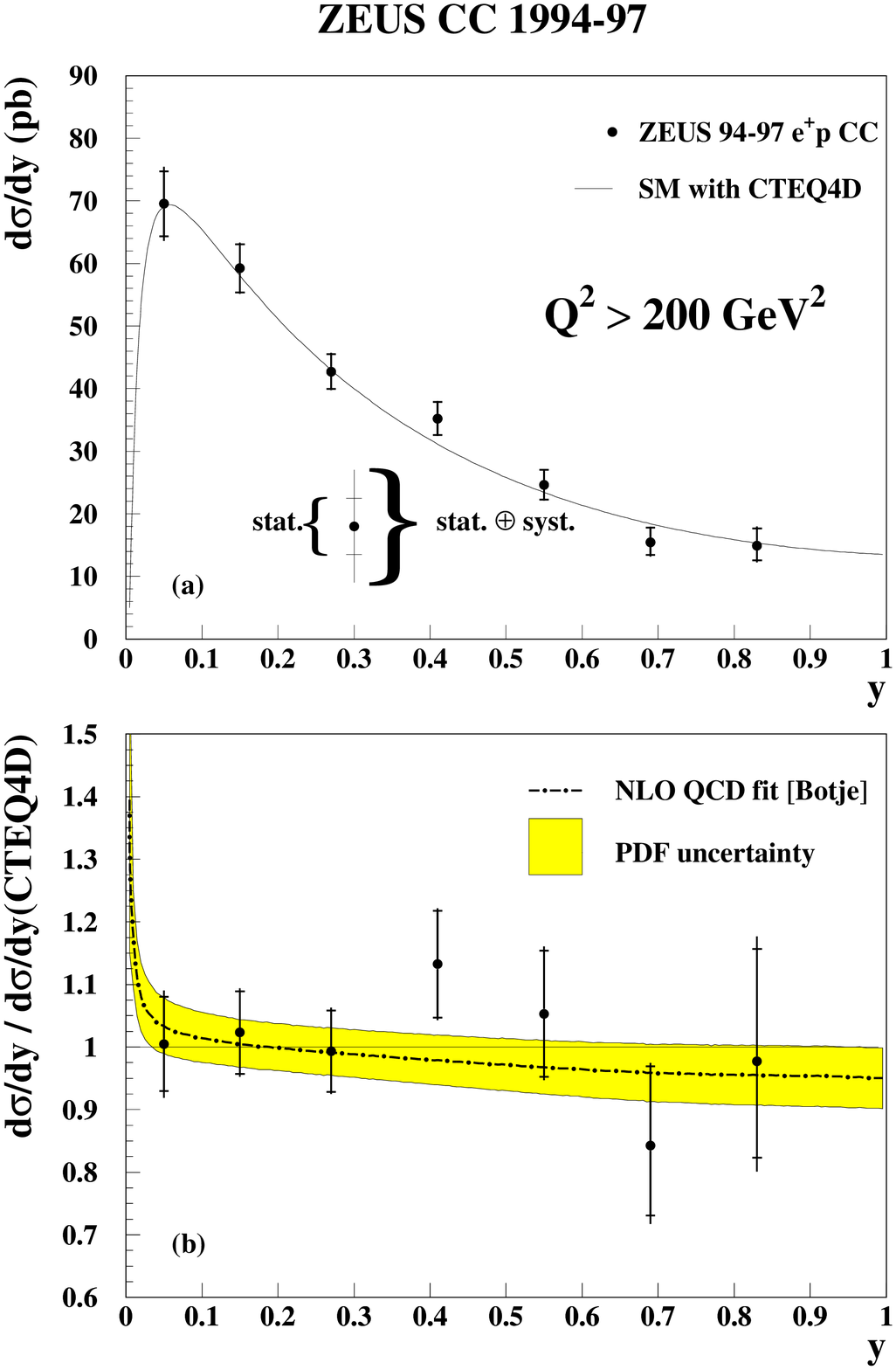}
  \end{center}
  \caption{
    (a) The $e^+p$ CC DIS Born cross section $d\sigma/dy$ for data
    (solid points) and the Standard Model (SM) expectation evaluated using
    the CTEQ4D PDFs.
    (b) The ratio of the measured cross section $d\sigma/dy$ to the
    SM expectation evaluated using the CTEQ4D PDFs.
    The statistical errors are indicated by the inner error bars (delimited by
    horizontal lines), while the full
    error bars show the total error obtained by adding the statistical and 
    systematic contributions in quadrature.
    Also shown by a dot-dashed line is the result of the NLO QCD fit
    together with the associated PDF uncertainties (shaded band).
    }
  \label{f:dsdy}
\end{figure}

The differential cross sections $d\sigma/dQ^2$, $d\sigma/dx$
and $d\sigma/dy$ are shown in Figs.~\ref{f:dsdq}(a), \ref{f:dsdx}(a)
and \ref{f:dsdy}(a) and are compiled\footnote{Tables~\ref{tab:q2xycorr} and~\ref{tab:xq2corr_1} contain details of the systematic uncertainties that are correlated between cross-section bins.} in
Tables~\ref{tab:dsdq}\,--\,\ref{tab:dsdy}.
The uncertainty of the measured cross sections are dominated by
statistical errors and are typically 7\,--\,20\%.
The cross section $d\sigma/dQ^2$ falls by a factor of about 50000
between $Q^2 = 280$ and 30000 GeV$^2$.  As a function of $x$,
the cross section $d\sigma/dx$ is largest at small $x$, showing a gradual
decrease from $x = 10^{-2}$ to $2{\cdot}10^{-1}$, followed by a sharp drop
towards $x = 0.6$.  The cross section $d\sigma/dy$ decreases slowly
as a function of $y$ for $y>0.1$.

The Standard Model cross sections from (\ref{e:Born}) using CTEQ4D
are also shown in Figs.~\ref{f:dsdq}\,--\,\ref{f:dsdy},
together with the ratios of the measured to the SM cross section.
The Standard Model describes the data well, with the possible
exception of $d\sigma/dx$, where the measurement at $x\gtrsim0.3$
is somewhat above the CTEQ4D prediction.

Also shown are the cross-section predictions obtained from a NLO QCD
fit~\cite{botje}, together with their uncertainties.
The fit is made to the data from fixed-target experiments and NC DIS
measurements at HERA in the region $Q^2 < 5000$~GeV$^2$.
Neither the CC data from HERA nor the recent high-$Q^2$ NC data
from ZEUS~\cite{ZNCpaper} are included in the fit.
The prediction from the fit also describes the data well, and in 
particular describes
$d\sigma/dx$ at higher $x$ better than the CTEQ4D prediction\footnote{
The CTEQ5 PDFs~\cite{CTEQ5}, which have been made available recently,
predict higher $d\sigma/dx$ than CTEQ4D in the region $0.1 < x < 0.6$.
}.
The $e^+p$ CC DIS cross section is dominated by the $d$-quark contribution
at high $x$, as can be seen from (\ref{e:F2}) and (\ref{e:F3}).
The possibility of a larger $d/u$ ratio than previously assumed
has been of interest in recent years, for example
see~\cite{bodek-yang,melnit}.
Modification~\cite{bodek-yang} of PDFs with an additional
term ($\delta(d/u)$; see Sect.~\ref{s:SystErr})
yields $d\sigma/dx$ close to the NLO QCD fit
as shown in Fig.~\ref{f:dsdx}(b).
For comparison, the prediction of the MRST~\cite{MRST} PDFs is also shown
in Fig.~\ref{f:dsdx}(b).

\subsection{Double differential cross section}
\label{sss:Double}

\begin{figure}
  \begin{center}
    \includegraphics[width=.85\textwidth]{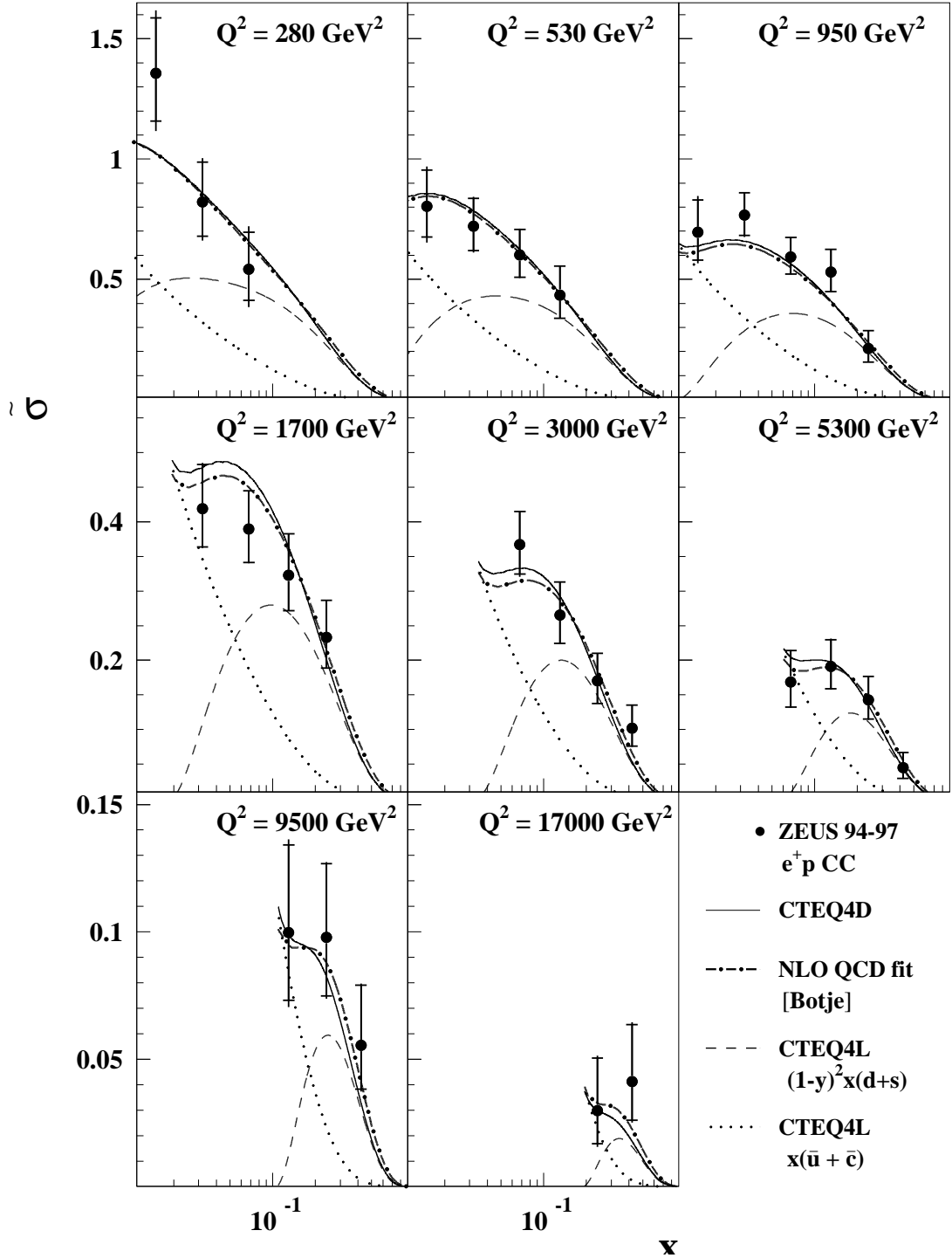}
  \end{center}
  \caption{
    The reduced cross section $\tilde{\sigma}$ as a function of $x$,
    for fixed values of $Q^2$.
    The dots represent the data, while the expectations of the
    Standard Model evaluated using the CTEQ4D PDFs are shown as the
    solid lines.
    For illustration, the leading-order PDF combinations $x(\bar{u}+\bar{c})$
    and $(1-y)^2 x(d+s)$, taken from the CTEQ4L parameterization, are plotted
    as dotted and dashed lines, respectively.
    Also shown is the result of the NLO QCD fit (dash-dotted line).
    }
  \label{f:d2sdxdq}
\end{figure}

\begin{figure}
  \begin{center}
    \includegraphics[width=.85\textwidth]{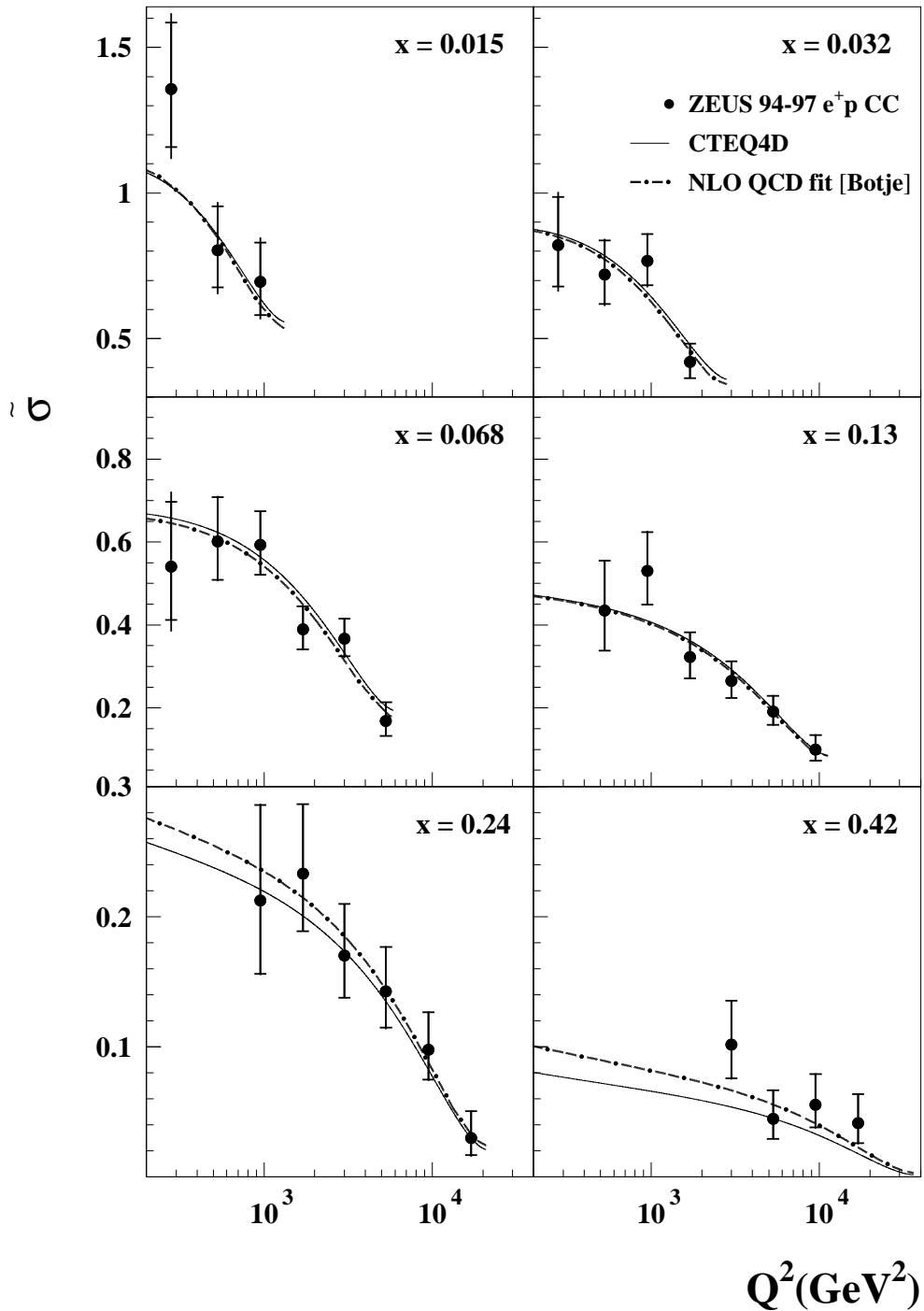}
  \end{center}
  \caption{
    The reduced cross section $\tilde{\sigma}$ as a function of $Q^2$,
    for fixed values of $x$.
    The dots represent the data, while the expectations of the
    Standard Model evaluated using the CTEQ4D PDFs are shown as the
    solid lines.
    Also shown is the result of the NLO QCD fit (dash-dotted line).
    }
  \label{f:d2sdxdq_1}
\end{figure}

The double differential cross section $d^2\sigma/dxdQ^2$ is compiled\footnotemark[1]
in Table~\ref{tab:ds2dxdq2_2}.
The reduced double differential cross section, $\tilde{\sigma}$, is
defined from the CC cross section as
\begin{equation}
  \tilde{\sigma} = \left[ 
   \frac{G^2_F}{2 \pi x} \Biggl( \frac{M^2_W}{M^2_W + Q^2}\Biggr)^2
                 \right]^{-1}
   \frac{d^2\sigma^{\rm CC}_{\rm Born}}{dx \, dQ^2}
   = \frac{1}{2}\left[Y_+ {\rm F}^{\rm CC}_2(x, Q^2) - Y_- x{\rm F}^{\rm CC}_3(x, Q^2) - y^2{\rm F}^{\rm CC}_L(x, Q^2) \right].
  \label{e:Red}
\end{equation}
The reduced cross sections as functions of $x$ and $Q^2$ are displayed
in Figs.~\ref{f:d2sdxdq} and \ref{f:d2sdxdq_1}.
The predictions of the CTEQ4D PDFs give a good description of the data,
although at the highest $x$ value the measured cross sections lie above
the predictions.
The predictions from the NLO QCD fit at high $x$ are higher than those
from CTEQ4D.

In leading-order QCD, $\tilde{\sigma}$
depends on the quark momentum distributions as follows
\begin{equation}
  \tilde{\sigma} = x\left[\bar{u} + \bar{c} + (1-y)^2 (d + s) \right].
  \label{e:LO}
\end{equation}
As a result, for fixed $Q^2$,
$\tilde{\sigma}$ at low $x$ (i.e. high $y$) is mainly
sensitive to the antiquark combination
$(\bar{u} + \bar{c})$ while at high $x$ (i.e. low $y$)
it is dominated by the quark combination $(d + s)$.
These PDF combinations evaluated with the leading-order CTEQ4L
parameterization are shown separately in Fig.~\ref{f:d2sdxdq}.
Both the quark and the antiquark combinations are required in order to
obtain a good description of the data.

\subsection{Comparison of NC and CC cross sections}
\label{sss:NCCC}

\begin{figure}
 \begin{center}
    \includegraphics[width=\textwidth]{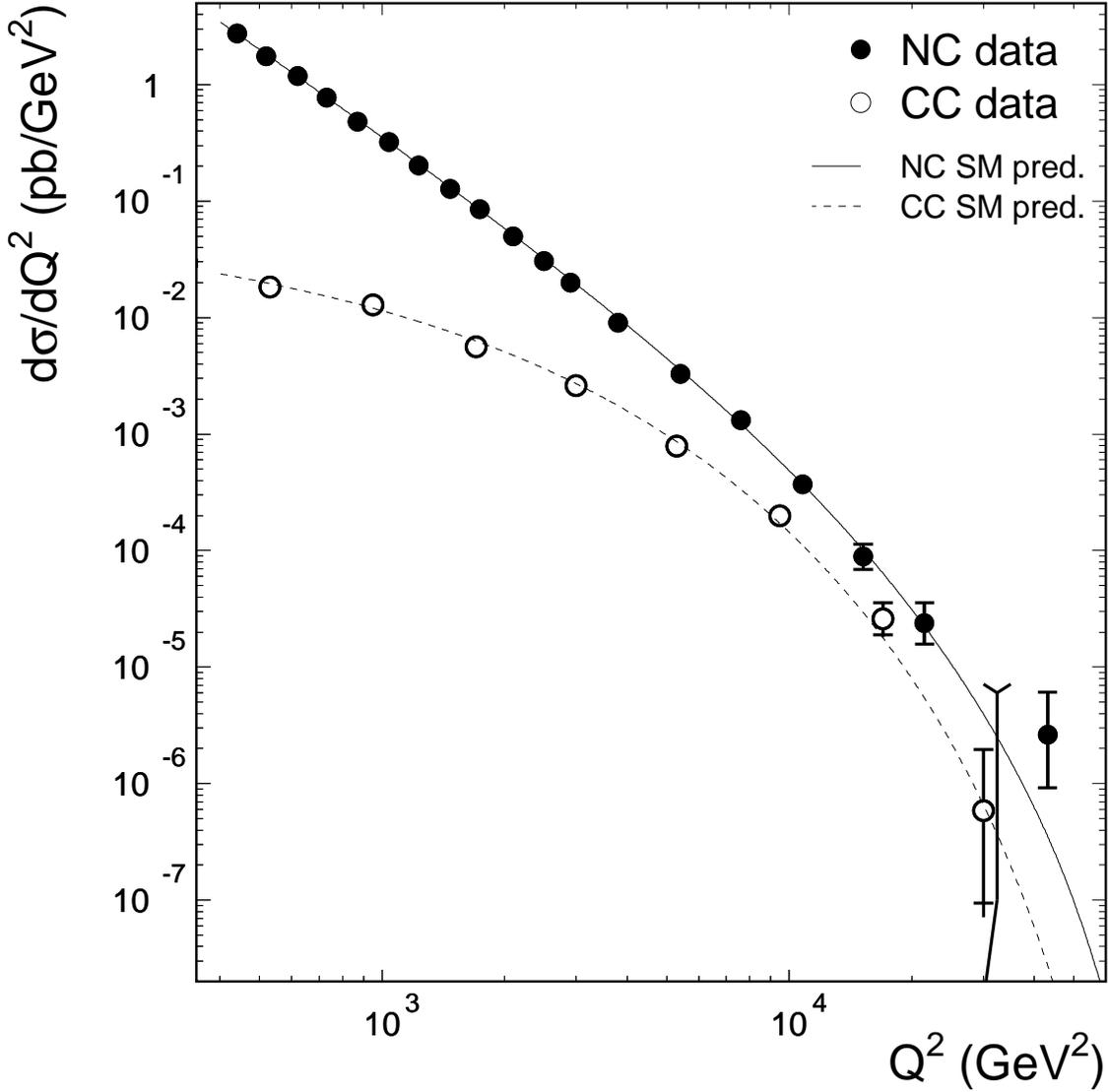}
  \end{center}
  \caption{
    Comparison of the differential cross sections $d\sigma/dQ^2$
    for NC (solid dots) and CC(open dots) deep inelastic $e^+p$ scattering
    from the ZEUS 1994-97 analysis.
    The lines represent the SM predictions evaluated using CTEQ4D PDFs.
    The second highest-$Q^2$ bin in the NC measurement contains no event: thus
    an upper limit of the cross section at 95\% confidence level is quoted
    and indicated by the arrow head.  This cross section is quoted at $Q^2 = 
    30400$~GeV$^2$ (very close to the CC point at $Q^2 = 30000$~GeV$^2$),
    as indicated by the lower end of the bent line.
    }
  \label{f:NCCC}
\end{figure}

Fig.~\ref{f:NCCC} compares the cross section $d\sigma/dQ^2$ for CC
scattering with the ZEUS result for NC scattering~\cite{ZNCpaper}.
At low $Q^2$, the CC cross section is much smaller than the NC cross section
due to the relative strength of the weak force compared to the
electromagnetic force.
However, the CC cross section decreases with $Q^2$ less rapidly than
that for NC scattering, reflecting the behavior of the $W$ propagator
as contrasted to the photon propagator which dominates NC scattering.
At $Q^2 \sim M_W^2, M_Z^2$, the CC and NC cross sections become
comparable; at these large $Q^2$, the weak and electromagnetic forces
are of similar strengths.
At yet higher $Q^2$, the rapid fall of both CC and NC cross sections
with $Q^2$ is due to the effects of the $W$ and $Z$ propagators,
the decrease of the parton densities with increasing $x$, and, in particular
for CC $e^+p$ scattering, the $(1-y)^2$ term in the cross section.
These observations were made in previous
HERA measurements~\cite{H1NCCC,ZEUSNCCC}; here they are clearly
demonstrated with a much improved precision.
%
%
\section{Electroweak analysis}
\label{sss:MW}

\begin{figure}
  \begin{center}
    \includegraphics[width=13.5cm]{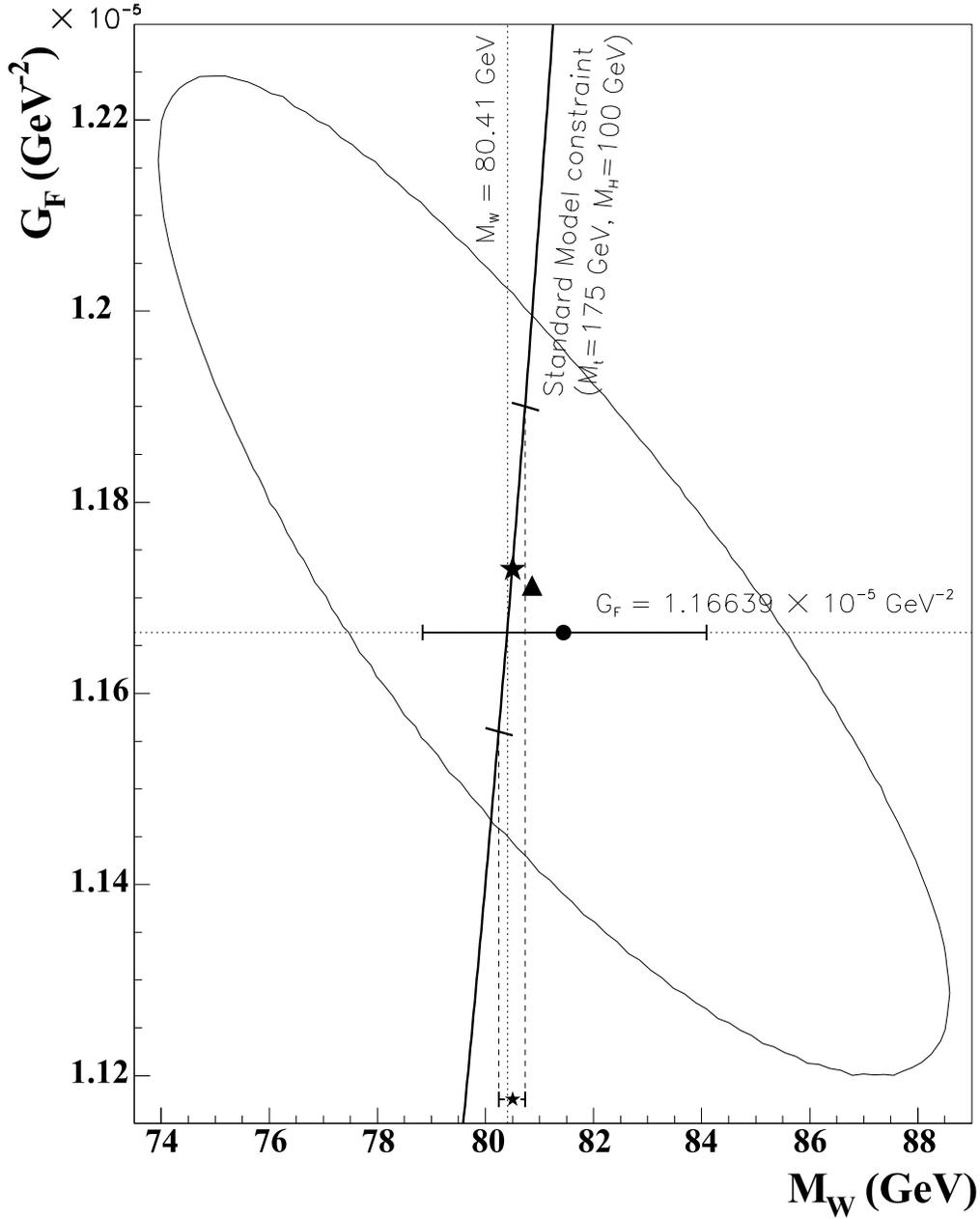}
  \end{center}
  \caption{
    The result of a fit of the CC DIS cross section to
    determine $G_F$ and $M_W$. 
    The triangle indicates the result of the fit, i.e. the
    position of $\chi^2_{\rm min}$, the minimum value of $\chi^2$. 
    The 70\% confidence level contour is shown as the ellipse.
    The dot with error bar shows the result of the
    `propagator-mass' fit, described in the text, in which
    the $\chi^2$ function is evaluated along the horizontal line
    $G_F=1.16639 \times 10^{-5}\,{\rm GeV^{-2}}$.
    The SM constraint implied by~(\ref{e:EWNLO}) is shown as the
    heavy solid line.
    The large star shows the position of $\chi^2_{\rm min}({\rm SM})$,
    the minimum of the $\chi^2$ function evaluated along the SM 
    constraint line.
    The solid bars crossing the SM constraint line show where
    $\chi^2({\rm SM}) = \chi^2_{\rm min}({\rm SM}) + 1$.
    The small star at the bottom with the error bar shows the value of $M_W$ 
    obtained in the `Standard Model fit'.
    Note that all errors and the confidence level contour
    correspond to statistical errors only.
    Also shown (dotted) are lines of constant $G_F$ 
    ($1.16639 \times 10^{-5}$~GeV$^{-2}$) and constant $M_W$ (80.41~GeV).
    }
  \label{f:GMUMW}
\end{figure}

The absolute magnitude of the CC cross section, described by
(\ref{e:Born}), is determined by the Fermi constant $G_F$ and the
PDFs, while the $Q^2$ dependence of the CC cross section includes the
propagator term $[M_W^2/(M_W^2+Q^2)]^2$, which produces substantial
damping of the cross section at high $Q^2$.
To compare the experimental results with the predictions of the Standard Model,
a $\chi^2$ fit to the measured differential cross section, $d\sigma/dQ^2$,
has been performed, treating $G_F$ and $M_W$ as free parameters
purely for the purpose of this exercise. The result of this fit is
\begin{eqnarray}
    G_F & = & \left( 
                     1.171      {\pm 0.034}       \,{\rm (stat.)} 
                               ^{+0.026}_{-0.032} \,{\rm (syst.)} 
                               ^{+0.016}_{-0.015} \,{\rm (PDF)} 
              \right)\times10^{-5}\,{\rm GeV^{-2}},
\end{eqnarray}
and
\begin{eqnarray}
    M_W & = &        80.8      ^{+4.9}_{-4.5} \,{\rm (stat.)} 
                               ^{+5.0}_{-4.3} \,{\rm (syst.)} 
                               ^{+1.4}_{-1.3} \,{\rm (PDF)~GeV}.
  \label{e:MWUncons}
\end{eqnarray}
The central values are obtained using the CTEQ4D PDFs.
The major sources of systematic uncertainty in the determination of the
cross section, namely the energy scale, the parton shower scheme and the
luminosity, are taken into account in the systematic errors.
The PDF errors quoted are obtained by re-evaluating the PDFs within the
uncertainties given by the NLO QCD fit~\cite{botje}.
The sensitivity of the result to variation of the value of $G_F$ assumed
in the extraction of the PDFs is negligible compared to the
uncertainties quoted above.
The point which gives minimum $\chi^2$ ($\chi^2_{\rm min}$) is displayed in
Fig.~\ref{f:GMUMW} as the triangle together with the 70\% confidence level
contour.  This contour was determined using statistical errors only.
The value of $G_F$ obtained is in good agreement with the value
$G_F = ( 1.16639 \pm 0.00001 ) \times 10^{-5}$~GeV$^{-2}$ obtained from
muon decay~\cite{PDG}, implying the universality of the CC interaction
over a wide range of $Q^2$.
The value of $M_W$ obtained agrees with the value of
\mbox{$M_W=80.41\pm 0.10$ GeV} from the PDG fit~\cite{PDG}, using 
time-like production of $W$ bosons at the Tevatron and at LEP.
Since CC DIS represents space-like exchange, the result (\ref{e:MWUncons}) is
complementary to measurements of $M_W$ from $p\bar{p}$ or $e^+e^-$
annihilation.  This result constitutes an important experimental
consistency check of the Standard Model.

Two more fits are performed to determine $M_W$ under more restrictive
theoretical assumptions.
First, by evaluating the $\chi^2$ function along the line
$G_F=1.16639 \times 10^{-5}\,{\rm GeV^{-2}}$, a measurement of the
`propagator mass' of the exchanged $W$ boson can be made.
The determination of the propagator mass is an important
test of the SM description of the CC in the space-like regime.
Second, a `Standard Model fit' may be performed by evaluating the
$\chi^2$ function along the SM constraint
\begin{equation}
  G_F = \frac{\pi \alpha}{\sqrt{2}} 
        \frac{M_Z^2}{\left(M_Z^2 - M_W^2 \right) M_W^2}
        \frac{1}{1 - \Delta r},
  \label{e:EWNLO}
\end{equation}
where $M_Z$ is the mass of the $Z$ boson and $\alpha$ is the fine
structure constant.
The term $\Delta r$ contains the radiative corrections to the lowest
order expression for $G_F$ and is a function of $\alpha$ and the
masses of the fundamental bosons and fermions~\cite{MWradcorr}.
The constraint implied by (\ref{e:EWNLO}) on $G_F$ and $M_W$
is also shown in Fig.~\ref{f:GMUMW} as the heavy solid line.
As can be seen from the plot, the value of $G_F$ has a strong
dependence on $M_W$.  
Therefore, within the context of the SM, the greatest sensitivity to
$M_W$ in this experiment may be obtained by combining the $M_W$
dependence of the propagator term in the CC cross section with the
$M_W$ dependence of $G_F$. 

The `propagator-mass' fit to the measured differential
cross section, $d\sigma / dQ^2$, with $G_F$ fixed to the value
$G_F=1.16639 \times 10^{-5}$~GeV$^{-2}$ yields the result
\begin{equation}
  M_W = 81.4^{+2.7}_{-2.6}{\rm (stat.)} {\pm 2.0} {\rm (syst.)}
                                   ^{+3.3}_{-3.0} {\rm (PDF)}~{\rm GeV},
\end{equation}
shown in Fig.~\ref{f:GMUMW} as the solid dot with horizontal error bars.
The use of the $G_F$ constraint has significantly reduced the
uncertainty on the estimation of $M_W$ compared to the value obtained
in the unconstrained fit (\ref{e:MWUncons}).  The result is in agreement with
the value of $M_W$ obtained by direct measurement,
$M_W=80.41\pm 0.10$ GeV~\cite{PDG}.

In order to use the SM constraint (\ref{e:EWNLO}),
$\alpha$, $M_Z$, and all fermion masses, other than the mass of the
top quark, $M_t$, are set to the PDG values~\cite{PDG}.
The central result of the fit was obtained with $M_t = 175$~GeV and 
the mass of the Higgs boson $M_H = 100$~GeV.
The $\chi^2$ function is evaluated along the line given by
the SM constraint and the position of the minimum, shown in
Fig.~\ref{f:GMUMW} as the large star, gives the `Standard Model fit'
result\footnote{It should be clearly
understood that (\ref{e:SMMW}) represents a constrained fit using assumptions
on the validity of the Standard Model which implies that the result
cannot be used in a global average of experimental values of $M_W$.}
\begin{equation}
  M_W = 80.50^{+0.24}_{-0.25} {\rm (stat.)} ^{+0.13}_{-0.16} {\rm (syst.)}
                                             {\pm 0.31}      {\rm (PDF)}
                                            ^{+0.03}_{-0.06} {\rm ( \Delta M_t, \Delta M_H, \Delta M_Z ) ~{\rm GeV}. }
  \label{e:SMMW}
\end{equation}
The error labeled $\Delta M_t, \Delta M_H, \Delta M_Z$ is obtained
by re-evaluating $M_W$ with $M_t$
in the range $170 < M_t < 180$~GeV, $M_H$ in the range 
$100 < M_H < 220$~GeV and $M_Z$ in the range $91.180 < M_Z <91.194$~GeV.
The dependence of $M_W$ on these changes is small, and the resulting
error is negligible compared with the other errors quoted above. 
This result is in agreement with the value of $M_W=80.35\pm 0.21$~GeV
obtained in fixed-target neutrino-nucleon DIS~\cite{NUTEV}.
The good agreement with both direct and indirect
determinations of $M_W$ indicates that the SM gives a consistent
description of a variety of phenomena over a wide range
of energy scales.
%
%
\section{Summary}
\label{s:conclusions}

Charged-current deep inelastic scattering,
$e^+ p \rightarrow \bar{\nu}_e X$, has been measured with the ZEUS detector
at HERA using \lumi~of data collected during 1994\,--\,1997.
Single differential cross sections $d\sigma/dQ^2$, $d\sigma/dx$ and
$d\sigma/dy$ have been presented with typical uncertainties of 7\,--\,20\%.
The cross section $d\sigma/dQ^2$ falls by a factor of about 50000
as $Q^2$ increases from 280 to 30000~GeV$^2$.
The double differential cross section $d^2\sigma/dxdQ^2$ has also been
measured.
A comparison between the data and Standard Model (SM) predictions
shows clearly that contributions from antiquarks ($\bar{u}$ and $\bar{c}$)
and quarks ($d$ and $s$) are both required by the data.
The predictions of the SM, using recent parton distribution functions,
give a good description of the full body of the data presented in this paper.

The charged-current (CC) cross-section results for $d\sigma/dQ^2$
presented here have been compared with the recent ZEUS results for
neutral-current (NC) scattering. The CC cross section is found to fall with
$Q^2$ less rapidly and to approach the NC cross section at
$Q^2 \gtrsim M_W^2, M_Z^2$, in agreement with previous observations at HERA.
This shows that the weak and electromagnetic 
forces reach similar strengths for $Q^2$ above $M^2_W, M^2_Z$.

   A fit to the data for $d\sigma/dQ^2$ with the Fermi constant $G_F$
and $M_W$ as independent parameters yields
\begin{eqnarray*}
    G_F & = & \left( 
                     1.171      {\pm 0.034}       \,{\rm (stat.)} 
                               ^{+0.026}_{-0.032} \,{\rm (syst.)} 
                               ^{+0.016}_{-0.015} \,{\rm (PDF)} 
              \right)\times10^{-5}\,{\rm GeV^{-2}},\\
    M_W & = &        80.8      ^{+4.9}_{-4.5} \,{\rm (stat.)} 
                               ^{+5.0}_{-4.3} \,{\rm (syst.)} 
                               ^{+1.4}_{-1.3} \,{\rm (PDF)~GeV}.
\end{eqnarray*}
A propagator fit, with $G_F$ fixed to the PDG value~\cite{PDG}, yields
\[ M_W = 81.4^{+2.7}_{-2.6}{\rm (stat.)} {\pm 2.0} {\rm (syst.)}
^{+3.3}_{-3.0} {\rm (PDF)}~{\rm GeV}. \]
These results show that the SM gives a consistent description
of charged-current induced time-like and space-like processes over
a wide range of virtualities.
%
%
\section*{Acknowledgements}

We appreciate the contributions to the construction and maintenance
of the ZEUS detector of the many people who are not listed as authors.
The HERA machine group and the DESY computing staff are especially
acknowledged for their success in providing excellent operation of the
collider and the data analysis environment.
We thank the DESY directorate for their strong support and encouragement.
%
%

%
%
\clearpage
%
%
\begin{table}
\caption
{
The differential cross
section $d\sigma/dQ^2$ for the reaction $e^{+} p \rightarrow \bar{\nu}_e X$.
The following quantities are given for each bin:
the $Q^2$ range; the value at which the cross section is quoted, $Q^2_c$;
the number of selected events, $N_{\rm obs}$;
the number of expected background events, $N_{\rm bg}$;
the acceptance, $\cal A$;
the radiative correction factor, ${\cal C}_{\rm rad}$
(see Sect.~\ref{s:Xsect});
the measured Born--level cross section, $d\sigma/dQ^2$;
and the Born--level cross 
section predicted by the Standard Model, using
CTEQ4D PDFs.
The first error of each measured cross section value gives the statistical
error, the second the systematic uncertainty.
}
\label{tab:dsdq}
\vskip 0.5 cm
\newcommand{\lw}[1]{\smash{\lower1.8ex\hbox{#1}}}\begin{center}
\begin{tabular}{|r@{ -- }r|r|r|r|c|c|l@{$\,$}l@{$\,$}l@{$\,$}l|l|} \hline
\multicolumn{2}{|c|}{$Q^2$ range} & \multicolumn{1}{c|}{$Q^2_{c}$} & \multicolumn{1}{c|}{\lw{$N_{obs}$}} & \multicolumn{1}{c|}{\lw{$N_{bg}$}} &
\lw{$\cal A$} & \lw{${\cal C}_{rad}$} & \multicolumn{5}{c|}{$d\sigma/dQ^2$(pb/GeV$^2$)}\\ \cline{8-12}
\multicolumn{2}{|c|}{(GeV$^2$)} & \multicolumn{1}{c|}{(GeV$^2$)} & & & & & \multicolumn{4}{c|}{measured} & \multicolumn{1}{c|}{SM}\\ \hline
  200 &   400 &   280 & 141 & 16.5 & 0.46 & 1.01 & $2.94$ & $\pm0.28$ & $^{+ 0.35}_{- 0.34}$ & $\cdot10^{-2}$& $2.80\cdot10^{-2}$\\
  400 &   711 &   530 & 173 & 5.3 & 0.64 & 1.01 & $1.82$ & $\pm0.14$ & $\pm 0.08$ & $\cdot10^{-2}$& $1.87\cdot10^{-2}$\\
  711 &  1265 &   950 & 248 & 2.3 & 0.74 & 1.01 & $1.29$ & $\pm0.08$ & $\pm 0.03$ & $\cdot10^{-2}$& $1.15\cdot10^{-2}$\\
 1265 &  2249 &  1700 & 205 & 2.3 & 0.79 & 1.03 & $5.62$ & $\pm0.40$ & $\pm 0.08$ & $\cdot10^{-3}$& $6.07\cdot10^{-3}$\\
 2249 &  4000 &  3000 & 169 & 2.4 & 0.81 & 1.04 & $2.62$ & $\pm0.20$ & $^{+ 0.04}_{- 0.09}$ & $\cdot10^{-3}$& $2.61\cdot10^{-3}$\\
 4000 &  7113 &  5300 &  91 & 0.6 & 0.82 & 1.05 & $7.91$ & $^{+ 0.93}_{- 0.83}$ & $^{+ 0.38}_{- 0.31}$ & $\cdot10^{-4}$& $8.29\cdot10^{-4}$\\
 7113 & 12649 &  9500 &  45 & 0.7 & 0.86 & 1.07 & $2.00$ & $^{+ 0.35}_{- 0.30}$ & $\pm 0.17$ & $\cdot10^{-4}$& $1.65\cdot10^{-4}$\\
12649 & 22494 & 17000 &  13 & 0.1 & 0.96 & 1.07 & $2.61$ & $^{+ 0.95}_{- 0.72}$ & $^{+ 0.45}_{- 0.38}$ & $\cdot10^{-5}$& $1.71\cdot10^{-5}$\\
22494 & 60000 & 30000 &   1 & 0.01 & 1.48 & 1.09 & $5.9$ & $^{+ 14.}_{- 4.9}$ & $^{+ 1.8}_{- 1.5}$ & $\cdot10^{-7}$& $6.24\cdot10^{-7}$\\ \hline
\end{tabular}
\end{center}
\end{table}
%
%
\clearpage
\begin{table}
\caption
{
The differential cross
section $d\sigma/dx$ for the reaction $e^{+} p \rightarrow \bar{\nu}_e X$
for $Q^2> 200$~GeV$^2$.
The following quantities are given for each bin:
the $x$ range; the value at which the cross section is quoted, $x_c$;
the number of selected events, $N_{\rm obs}$;
the number of expected background events, $N_{\rm bg}$;
the acceptance, $\cal A$;
the radiative correction factor, ${\cal C}_{\rm rad}$
(see Sect.~\ref{s:Xsect});
the measured Born--level cross section, $d\sigma/dx$;
and the Born--level cross 
section predicted by the Standard Model, using
CTEQ4D PDFs.
The first error of each measured cross section value gives the statistical
error, the second the systematic uncertainty.
}
\label{tab:dsdx}
\vskip 0.5 cm
\newcommand{\lw}[1]{\smash{\lower1.8ex\hbox{#1}}}\begin{center}
\begin{tabular}{|r@{ -- }r|r|r|r|c|c|l@{$\,$}l@{$\,$}l@{$\,$}l|l|} \hline
\multicolumn{2}{|c|}{\lw{$x$ range}} & \multicolumn{1}{c|}{\lw{$x_{c}$}} & \multicolumn{1}{c|}{\lw{$N_{obs}$}} & \multicolumn{1}{c|}{\lw{$N_{bg}$}} &
\lw{$\cal A$} & \lw{${\cal C}_{rad}$} & \multicolumn{5}{c|}{$d\sigma/dx$(pb)}\\ \cline{8-12}
\multicolumn{2}{|c|}{} & & & & && \multicolumn{4}{c|}{measured}& \multicolumn{1}{c|}{SM}\\ \hline
0.0100 & 0.0215 &  0.015 & 136 &   4.0 & 0.56 & 1.04 & $4.50$ & $\pm0.40$ & $^{+ 0.34}_{- 0.35}$ & $\cdot10^{2}$& $3.97\cdot10^{2}$\\
0.0215 & 0.0464 &  0.032 & 246 &   7.4 & 0.79 & 1.02 & $2.64$ & $\pm0.17$ & $\pm 0.07$ & $\cdot10^{2}$& $2.76\cdot10^{2}$\\
0.0464 & 0.1000 &  0.068 & 306 &  10.0 & 0.85 & 1.02 & $1.44$ & $\pm0.08$ & $\pm 0.03$ & $\cdot10^{2}$& $1.50\cdot10^{2}$\\
0.1000 & 0.1780 &  0.130 & 200 &   2.9 & 0.82 & 1.01 & $6.88$ & $\pm0.49$ & $^{+ 0.17}_{- 0.12}$ & $\cdot10^{1}$& $7.04\cdot10^{1}$\\
0.1780 & 0.3160 &  0.240 & 124 &   0.6 & 0.73 & 1.01 & $2.57$ & $\pm0.23$ & $^{+ 0.09}_{- 0.10}$ & $\cdot10^{1}$& $2.39\cdot10^{1}$\\
0.3160 & 0.5620 &  0.420 &  46 &  0.06 & 0.56 & 1.00 & $6.8$ & $^{+ 1.2}_{- 1.0}$ & $\pm 0.6$ & $\cdot10^{0}$& $4.56\cdot10^{0}$\\
0.5620 & 1.0000 &  0.650 &   3 &  0.00 & 0.36 & 1.02 & $8.1$ & $^{+ 7.8}_{- 4.4}$ & $^{+ 2.1}_{- 1.6}$ & $\cdot10^{-1}$& $3.55\cdot10^{-1}$\\ \hline
\end{tabular}
\end{center}
\end{table}
%
%
\clearpage
\begin{table}
\caption
{
The differential cross
section $d\sigma/dy$ for the reaction $e^{+} p \rightarrow \bar{\nu}_e X$
for $Q^2> 200$~GeV$^2$.
The following quantities are given for each bin:
the $y$ range; the value at which the cross section is quoted, $y_c$;
the number of selected events, $N_{\rm obs}$;
the number of expected background events, $N_{\rm bg}$;
the acceptance, $\cal A$;
the radiative correction factor, ${\cal C}_{\rm rad}$
(see Sect.~\ref{s:Xsect});
the measured Born--level cross section, $d\sigma/dy$;
and the Born--level cross 
section predicted by the Standard Model, using
CTEQ4D PDFs.
The first error of each measured cross section value gives the statistical
error, the second the systematic uncertainty.
}
\vskip 0.5 cm
\label{tab:dsdy}
\newcommand{\lw}[1]{\smash{\lower1.8ex\hbox{#1}}}\begin{center}
\begin{tabular}{|r@{ -- }r|r|r|r|c|c|l@{$\,$}l@{$\,$}l@{$\,$}l|l|} \hline
\multicolumn{2}{|c|}{\lw{$y$ range}} & \multicolumn{1}{c|}{\lw{$y_{c}$}} & \multicolumn{1}{c|}{\lw{$N_{obs}$}} & \multicolumn{1}{c|}{\lw{$N_{bg}$}} &
\lw{$\cal A$} & \lw{${\cal C}_{rad}$} & \multicolumn{5}{c|}{$d\sigma/dy$(pb)}\\ \cline{8-12}
\multicolumn{2}{|c|}{} & & & & & & \multicolumn{4}{c|}{measured} & \multicolumn{1}{c|}{SM}\\ \hline
  0.00 &   0.10 &   0.05 & 192 & 7.3 & 0.64 & 0.98 & $6.95$ & $\pm0.52$ & $\pm 0.28$ & $\cdot10^{1}$& $6.92\cdot10^{1}$\\
  0.10 &   0.20 &   0.15 & 249 & 5.2 & 0.85 & 0.99 & $5.92$ & $\pm0.38$ & $\pm 0.14$ & $\cdot10^{1}$& $5.79\cdot10^{1}$\\
  0.20 &   0.34 &   0.27 & 240 & 2.6 & 0.83 & 1.01 & $4.27$ & $\pm0.28$ & $^{+ 0.11}_{- 0.09}$ & $\cdot10^{1}$& $4.30\cdot10^{1}$\\
  0.34 &   0.48 &   0.41 & 185 & 5.0 & 0.78 & 1.03 & $3.52$ & $\pm0.27$ & $\pm 0.08$ & $\cdot10^{1}$& $3.11\cdot10^{1}$\\
  0.48 &   0.62 &   0.55 & 117 & 3.8 & 0.70 & 1.05 & $2.46$ & $\pm0.24$ & $^{+ 0.09}_{- 0.08}$ & $\cdot10^{1}$& $2.34\cdot10^{1}$\\
  0.62 &   0.76 &   0.69 &  61 & 2.1 & 0.60 & 1.07 & $1.55$ & $^{+ 0.23}_{- 0.20}$ & $^{+ 0.07}_{- 0.10}$ & $\cdot10^{1}$& $1.84\cdot10^{1}$\\
  0.76 &   0.90 &   0.83 &  42 & 1.1 & 0.44 & 1.10 & $1.49$ & $^{+ 0.27}_{- 0.24}$ & $\pm 0.13$ & $\cdot10^{1}$& $1.53\cdot10^{1}$\\ \hline
\end{tabular}
\end{center}
\end{table}
%
%
\clearpage
\begin{table}
\rotatebox{90}{
\begin{minipage}[b]{\textheight}
\caption
{
The double differential cross
section $d\sigma/dxdQ^2$
for the reaction $e^{+} p \rightarrow \bar{\nu}_e X$.
The following quantities are given for each bin:
the $x$ and $Q^2$ range;
the values at which the cross section is quoted, $x_c$ and $Q^2_c$;
the number of selected events, $N_{\rm obs}$;
the number of expected background events, $N_{\rm bg}$;
the acceptance, $\cal A$;
the radiative correction factor, ${\cal C}_{\rm rad}$
(see Sect.~\ref{s:Xsect});
the measured Born--level cross section, $d\sigma/dxdQ^2$;
and the Born--level cross 
section predicted by the Standard Model, using
CTEQ4D PDFs.
The first error of each measured cross section value gives the statistical
error, the second the systematic uncertainty.
}
\label{tab:ds2dxdq2_1}
\vskip 0.5 cm
\newcommand{\lw}[1]{\smash{\lower1.8ex\hbox{#1}}}\begin{center}
\begin{tabular}{|r@{ -- }r|r@{ -- }r|r|r|r|r|c|c|l@{$\,$}l@{$\,$}l@{$\,$}l|l|} \hline
\multicolumn{2}{|c|}{$Q^2$ range} & \multicolumn{2}{|c|}{$x$ range} & \multicolumn{1}{c|}{$Q^2_{c}$} & \multicolumn{1}{c|}{\lw{$x_{c}$}} & \multicolumn{1}{c|}{\lw{$N_{obs}$}} & \multicolumn{1}{c|}{\lw{$N_{bg}$}} &
\lw{$\cal A$} & \lw{${\cal C}_{rad}$} & \multicolumn{5}{c|}{$d^2\sigma/dxdQ^2$(pb/GeV$^2$)}\\ \cline{11-15}
\multicolumn{2}{|c|}{(GeV$^2$)} & \multicolumn{2}{|c|}{} & \multicolumn{1}{c|}{(GeV$^2$)} & & & & & & \multicolumn{4}{c|}{measured} & \multicolumn{1}{c|}{SM}\\ \hline
   200 &    400 & 0.0100 & 0.0215 &    280 & 0.015 &  50 & 2.0 & 0.64 & 1.00 & $7.0$ & $^{+ 1.2}_{- 1.0}$ & $\pm 0.7$ & $\cdot10^{-1}$& $5.27\cdot10^{-1}$\\
\multicolumn{2}{|c|}{} & 0.0215 & 0.0464 &  & 0.032 &  42 & 4.7 & 0.80 & 0.99 & $1.99$ & $^{+ 0.40}_{- 0.34}$ & $^{+ 0.19}_{- 0.18}$ & $\cdot10^{-1}$& $2.07\cdot10^{-1}$\\
\multicolumn{2}{|c|}{} & 0.0464 & 0.1000 &  & 0.068 &  28 & 5.9 & 0.71 & 0.97 & $6.2$ & $^{+ 1.8}_{- 1.5}$ & $\pm 1.0$ & $\cdot10^{-2}$& $7.49\cdot10^{-2}$\\
   400 &    711 & 0.0100 & 0.0215 &    530 & 0.015 &  42 & 1.6 & 0.64 & 1.02 & $3.86$ & $^{+ 0.72}_{- 0.62}$ & $^{+ 0.33}_{- 0.39}$ & $\cdot10^{-1}$& $4.10\cdot10^{-1}$\\
\multicolumn{2}{|c|}{} & 0.0215 & 0.0464 &  & 0.032 &  52 & 0.8 & 0.87 & 0.99 & $1.62$ & $^{+ 0.26}_{- 0.23}$ & $\pm 0.09$ & $\cdot10^{-1}$& $1.76\cdot10^{-1}$\\
\multicolumn{2}{|c|}{} & 0.0464 & 0.1000 &  & 0.068 &  44 & 1.0 & 0.86 & 0.98 & $6.4$ & $^{+ 1.1}_{- 1.0}$ & $^{+ 0.3}_{- 0.2}$ & $\cdot10^{-2}$& $6.60\cdot10^{-2}$\\
\multicolumn{2}{|c|}{} & 0.1000 & 0.1780 &  & 0.130 &  21 & 0.5 & 0.77 & 0.98 & $2.41$ & $^{+ 0.67}_{- 0.53}$ & $^{+ 0.13}_{- 0.07}$ & $\cdot10^{-2}$& $2.44\cdot10^{-2}$\\
   711 &   1265 & 0.0100 & 0.0215 &    950 & 0.015 &  37 & 0.3 & 0.50 & 1.06 & $2.97$ & $^{+ 0.58}_{- 0.49}$ & $^{+ 0.30}_{- 0.25}$ & $\cdot10^{-1}$& $2.72\cdot10^{-1}$\\
\multicolumn{2}{|c|}{} & 0.0215 & 0.0464 &  & 0.032 &  85 & 0.4 & 0.88 & 1.01 & $1.53$ & $^{+ 0.19}_{- 0.17}$ & $^{+ 0.04}_{- 0.07}$ & $\cdot10^{-1}$& $1.32\cdot10^{-1}$\\
\multicolumn{2}{|c|}{} & 0.0464 & 0.1000 &  & 0.068 &  69 & 0.5 & 0.89 & 0.99 & $5.59$ & $^{+ 0.76}_{- 0.68}$ & $^{+ 0.12}_{- 0.11}$ & $\cdot10^{-2}$& $5.31\cdot10^{-2}$\\
\multicolumn{2}{|c|}{} & 0.1000 & 0.1780 &  & 0.130 &  43 & 0.2 & 0.84 & 0.98 & $2.61$ & $^{+ 0.46}_{- 0.40}$ & $\pm 0.10$ & $\cdot10^{-2}$& $2.02\cdot10^{-2}$\\
\multicolumn{2}{|c|}{} & 0.1780 & 0.3160 &  & 0.240 &  14 & 0.0 & 0.66 & 0.99 & $5.7$ & $^{+ 2.0}_{- 1.5}$ & $\pm 0.3$ & $\cdot10^{-3}$& $5.90\cdot10^{-3}$\\
  1265 &   2249 & 0.0215 & 0.0464 &   1700 & 0.032 &  58 & 0.9 & 0.77 & 1.05 & $6.9$ & $^{+ 1.1}_{- 0.9}$ & $^{+ 0.3}_{- 0.2}$ & $\cdot10^{-2}$& $7.90\cdot10^{-2}$\\
\multicolumn{2}{|c|}{} & 0.0464 & 0.1000 &  & 0.068 &  66 & 1.1 & 0.91 & 1.02 & $3.03$ & $^{+ 0.43}_{- 0.38}$ & $^{+ 0.05}_{- 0.07}$ & $\cdot10^{-2}$& $3.63\cdot10^{-2}$\\
\multicolumn{2}{|c|}{} & 0.1000 & 0.1780 &  & 0.130 &  40 & 0.3 & 0.89 & 1.00 & $1.31$ & $^{+ 0.24}_{- 0.21}$ & $^{+ 0.01}_{- 0.03}$ & $\cdot10^{-2}$& $1.47\cdot10^{-2}$\\
\multicolumn{2}{|c|}{} & 0.1780 & 0.3160 &  & 0.240 &  28 & 0.3 & 0.83 & 1.00 & $5.1$ & $^{+ 1.2}_{- 1.0}$ & $\pm 0.1$ & $\cdot10^{-3}$& $4.42\cdot10^{-3}$\\ \hline
\end{tabular}
\end{center}
\end{minipage}
}
\end{table}
\clearpage
\addtocounter{table}{-1}
\begin{table}
\rotatebox{90}{
\begin{minipage}[b]{\textheight}
\caption
{
continued.
}
\label{tab:ds2dxdq2_2}
\vskip 0.5 cm
\newcommand{\lw}[1]{\smash{\lower1.8ex\hbox{#1}}}\begin{center}
\begin{tabular}{|r@{ -- }r|r@{ -- }r|r|r|r|r|c|c|l@{$\,$}l@{$\,$}l@{$\,$}l|l|} \hline
\multicolumn{2}{|c|}{$Q^2$ range} & \multicolumn{2}{|c|}{$x$ range} & \multicolumn{1}{c|}{$Q^2_{c}$} & \multicolumn{1}{c|}{\lw{$x_{c}$}} & \multicolumn{1}{c|}{\lw{$N_{obs}$}} & \multicolumn{1}{c|}{\lw{$N_{bg}$}} &
\lw{$\cal A$} & \lw{${\cal C}_{rad}$} & \multicolumn{5}{c|}{$d^2\sigma/dxdQ^2$(pb/GeV$^2$)}\\ \cline{11-15}
\multicolumn{2}{|c|}{(GeV$^2$)} & \multicolumn{2}{|c|}{} & \multicolumn{1}{c|}{(GeV$^2$)} & & & & & & \multicolumn{4}{c|}{measured} & \multicolumn{1}{c|}{SM}\\ \hline
  2249 &   4000 & 0.0464 & 0.1000 &   3000 & 0.068 &  76 & 1.1 & 0.88 & 1.06 & $2.12$ & $^{+ 0.28}_{- 0.25}$ & $^{+ 0.02}_{- 0.11}$ & $\cdot10^{-2}$& $1.92\cdot10^{-2}$\\
\multicolumn{2}{|c|}{} & 0.1000 & 0.1780 &  & 0.130 &  42 & 0.2 & 0.89 & 1.03 & $8.0$ & $^{+ 1.4}_{- 1.2}$ & $^{+ 0.2}_{- 0.1}$ & $\cdot10^{-3}$& $8.86\cdot10^{-3}$\\
\multicolumn{2}{|c|}{} & 0.1780 & 0.3160 &  & 0.240 &  27 & 0.1 & 0.86 & 1.01 & $2.79$ & $^{+ 0.65}_{- 0.54}$ & $^{+ 0.08}_{- 0.09}$ & $\cdot10^{-3}$& $2.84\cdot10^{-3}$\\
\multicolumn{2}{|c|}{} & 0.3160 & 0.5620 &  & 0.420 &  15 & 0.0 & 0.75 & 1.00 & $9.5$ & $^{+ 3.2}_{- 2.4}$ & $^{+ 0.4}_{- 0.6}$ & $\cdot10^{-4}$& $5.03\cdot10^{-4}$\\
  4000 &   7113 & 0.0464 & 0.1000 &   5300 & 0.068 &  22 & 0.4 & 0.69 & 1.07 & $6.3$ & $^{+ 1.7}_{- 1.4}$ & $\pm 0.4$ & $\cdot10^{-3}$& $7.65\cdot10^{-3}$\\
\multicolumn{2}{|c|}{} & 0.1000 & 0.1780 &  & 0.130 &  35 & 0.2 & 0.92 & 1.04 & $3.75$ & $^{+ 0.75}_{- 0.63}$ & $^{+ 0.23}_{- 0.19}$ & $\cdot10^{-3}$& $3.86\cdot10^{-3}$\\
\multicolumn{2}{|c|}{} & 0.1780 & 0.3160 &  & 0.240 &  26 & 0.1 & 0.89 & 1.03 & $1.51$ & $^{+ 0.36}_{- 0.30}$ & $^{+ 0.07}_{- 0.09}$ & $\cdot10^{-3}$& $1.43\cdot10^{-3}$\\
\multicolumn{2}{|c|}{} & 0.3160 & 0.5620 &  & 0.420 &   8 & 0.0 & 0.84 & 1.04 & $2.7$ & $^{+ 1.3}_{- 1.0}$ & $^{+ 0.3}_{- 0.2}$ & $\cdot10^{-4}$& $2.73\cdot10^{-4}$\\
  7113 &  12649 & 0.1000 & 0.1780 &   9500 & 0.130 &  15 & 0.6 & 0.84 & 1.08 & $1.06$ & $^{+ 0.37}_{- 0.28}$ & $\pm 0.13$ & $\cdot10^{-3}$& $1.05\cdot10^{-3}$\\
\multicolumn{2}{|c|}{} & 0.1780 & 0.3160 &  & 0.240 &  18 & 0.1 & 0.94 & 1.05 & $5.6$ & $^{+ 1.7}_{- 1.3}$ & $\pm 0.4$ & $\cdot10^{-4}$& $4.74\cdot10^{-4}$\\
\multicolumn{2}{|c|}{} & 0.3160 & 0.5620 &  & 0.420 &  10 & 0.0 & 0.90 & 1.05 & $1.82$ & $^{+ 0.78}_{- 0.57}$ & $\pm 0.19$ & $\cdot10^{-4}$& $1.09\cdot10^{-4}$\\
 12649 &  22494 & 0.1780 & 0.3160 &  17000 & 0.240 &   5 & 0.1 & 1.04 & 1.07 & $8.0$ & $^{+ 5.5}_{- 3.5}$ & $^{+ 1.7}_{- 1.5}$ & $\cdot10^{-5}$& $7.97\cdot10^{-5}$\\
\multicolumn{2}{|c|}{} & 0.3160 & 0.5620 &  & 0.420 &   7 & 0.0 & 1.03 & 1.06 & $6.3$ & $^{+ 3.4}_{- 2.3}$ & $^{+ 1.0}_{- 0.9}$ & $\cdot10^{-5}$& $2.72\cdot10^{-5}$\\ \hline
\end{tabular}
\end{center}
\end{minipage}
}
\end{table}
%
%
\begin{table}
\caption
{
The differential cross
sections $d\sigma/dQ^2$, $d\sigma/dx$ and $d\sigma/dy$
for the reaction $e^{+} p \rightarrow \bar{\nu}_e X$.
The following quantities are given for each bin:
the value at which the cross section is quoted;
the measured Born--level cross section;
the statistical uncertainty;
the total systematic uncertainty;
the uncorrelated systematic uncertainty and
those systematic uncertainties with significant (assumed 100\%)
correlations between cross-seciton bins.
The systematic uncertainties considered to be correlated were:
the FCAL energy scale ($\delta_1$);
the BCAL energy scale ($\delta_2$) and
the uncertainty in the parton-shower scheme ($\delta_3$).
}
\label{tab:q2xycorr}
\vskip 0.5 cm
\newcommand{\lw}[1]{\smash{\lower1.8ex\hbox{#1}}}\begin{center}
\begin{tabular}{|r|c|r|r||r|r|r|r|} \hline
\multicolumn{8}{|c|}{$d\sigma/dQ^2$}\\ \hline
 \multicolumn{1}{|c|}{$Q^2_c$}
 & \multicolumn{1}{c|}{$d\sigma/dQ^2$ (pb/GeV$^2$)}
 & \multicolumn{1}{c|}{$\delta_{\rm stat}$ (\%)}
 & \multicolumn{1}{c||}{$\delta_{\rm sys}$ (\%)}
 & \multicolumn{1}{c|}{$\delta_{\rm unc}$ (\%)}
 & \multicolumn{1}{c|}{$\delta_{\rm 1}$ (\%)}
 & \multicolumn{1}{c|}{$\delta_{\rm 2}$ (\%)}
 & \multicolumn{1}{c|}{$\delta_{\rm 3}$ (\%)}
\\ \hline
  280& $2.94\cdot10^{-2}$ & $\pm   9.5$         & $\pm 12.$          & $^{+ 8.4}_{- 8.3}$ & $^{- 1.0}_{+ 2.4}$ & $^{- 1.0}_{+ 1.9}$ & $^{- 7.9}_{+ 7.9}$\\
  530& $1.82\cdot10^{-2}$ & $\pm   7.8$         & $^{+ 4.2}_{- 4.5}$ & $^{+ 3.6}_{- 3.7}$ & $^{- 1.4}_{+ 1.0}$ & $^{- 1.2}_{+ 0.7}$ & $^{- 1.9}_{+ 1.9}$\\
  950& $1.29\cdot10^{-2}$ & $\pm   6.4$         & $^{+ 2.3}_{- 2.2}$ & $^{+ 2.0}_{- 1.8}$ & $^{+ 0.3}_{+ 0.5}$ & $^{- 1.0}_{+ 1.0}$ & $^{- 0.5}_{+ 0.5}$\\
 1700& $5.62\cdot10^{-3}$ & $\pm   7.1$         & $^{+ 1.5}_{- 1.4}$ & $^{+ 1.0}_{- 1.1}$ & $^{+ 0.4}_{- 0.2}$ & $^{+ 0.5}_{+ 0.0}$ & $^{+ 0.9}_{- 0.9}$\\
 3000& $2.62\cdot10^{-3}$ & $\pm   7.8$         & $^{+ 1.3}_{- 3.3}$ & $^{+ 0.8}_{- 1.2}$ & $^{+ 0.6}_{- 2.1}$ & $^{+ 0.4}_{- 2.2}$ & $^{- 0.8}_{+ 0.8}$\\
 5300& $7.91\cdot10^{-4}$ & $^{+ 12.}_{- 11.}$  & $^{+ 4.8}_{- 3.9}$ & $^{+ 0.8}_{- 0.6}$ & $^{+ 3.0}_{- 1.7}$ & $^{+ 3.6}_{- 3.4}$ & $^{+ 0.4}_{- 0.4}$\\
 9500& $2.00\cdot10^{-4}$ & $^{+ 18.}_{- 15.}$  & $\pm 8.6$          & $\pm 0.7$          & $^{+ 4.4}_{- 4.5}$ & $^{+ 7.1}_{- 7.0}$ & $^{+ 2.2}_{- 2.2}$\\
17000& $2.61\cdot10^{-5}$ & $^{+ 37.}_{- 28.}$  & $^{+ 17.}_{- 15.}$ & $^{+ 1.3}_{- 1.0}$ & $^{+ 7.1}_{- 5.6}$ & $^{+ 15.}_{- 13.}$ & $^{+ 3.3}_{- 3.3}$\\
30000& $5.9 \cdot10^{-7}$ & $^{+ 233.}_{- 84.}$ & $^{+ 30.}_{- 26.}$ & $^{+ 3.2}_{- 3.3}$ & $^{+ 9.0}_{- 9.4}$ & $^{+ 28.}_{- 23.}$ & $^{+ 7.6}_{- 7.6}$\\ \hline
%
%
\multicolumn{8}{|c|}{$d\sigma/dx$}\\ \hline
 \multicolumn{1}{|c|}{$x_c$}
 & \multicolumn{1}{c|}{$d\sigma/dx$ (pb)}
 & \multicolumn{1}{c|}{$\delta_{\rm stat}$ (\%)}
 & \multicolumn{1}{c||}{$\delta_{\rm sys}$ (\%)}
 & \multicolumn{1}{c|}{$\delta_{\rm unc}$ (\%)}
 & \multicolumn{1}{c|}{$\delta_{\rm 1}$ (\%)}
 & \multicolumn{1}{c|}{$\delta_{\rm 2}$ (\%)}
 & \multicolumn{1}{c|}{$\delta_{\rm 3}$ (\%)}
\\ \hline
 0.015& $4.50\cdot10^{2}$ & $\pm   8.8$        & $^{+ 7.6}_{- 7.9}$ & $\pm 6.1$          & $^{- 1.0}_{+ 1.5}$ & $^{- 2.3}_{+ 0.6}$ & $^{- 4.3}_{+ 4.3}$\\
 0.032& $2.64\cdot10^{2}$ & $\pm   6.6$        & $^{+ 2.7}_{- 2.6}$ & $^{+ 2.4}_{- 2.5}$ & $^{- 0.2}_{+ 0.2}$ & $^{- 0.3}_{+ 0.7}$ & $^{- 0.7}_{+ 0.7}$\\
 0.068& $1.44\cdot10^{2}$ & $\pm   5.9$        & $^{+ 1.9}_{- 2.2}$ & $\pm 1.5$          & $^{- 0.4}_{+ 0.5}$ & $^{+ 0.4}_{- 1.2}$ & $^{- 1.0}_{+ 1.0}$\\
 0.130& $6.88\cdot10^{1}$ & $\pm   7.2$        & $^{+ 2.4}_{- 1.8}$ & $\pm 1.2$          & $^{+ 0.4}_{- 0.3}$ & $^{+ 2.0}_{- 1.3}$ & $^{- 0.1}_{+ 0.1}$\\
 0.240& $2.57\cdot10^{1}$ & $\pm   9.0$        & $^{+ 3.5}_{- 3.7}$ & $\pm 1.0$          & $^{+ 2.8}_{- 3.2}$ & $^{+ 1.8}_{- 1.6}$ & $^{+ 0.2}_{- 0.2}$\\
 0.420& $6.8\cdot10^{0}$  & $^{+ 17.}_{- 15.}$ & $^{+ 8.3}_{- 8.4}$ & $\pm 2.4$          & $^{+ 7.8}_{- 7.9}$ & $^{+ 0.9}_{- 1.1}$ & $^{+ 1.5}_{- 1.5}$\\
 0.650& $8.1\cdot10^{-1}$ & $^{+ 97.}_{- 54.}$ & $^{+ 26.}_{- 20.}$ & $^{+ 5.2}_{- 5.1}$ & $^{+ 25.}_{- 19.}$ & $^{+ 2.1}_{- 0.9}$ & $^{+ 1.3}_{- 1.3}$\\ \hline
%
%
\multicolumn{8}{|c|}{$d\sigma/dy$}\\ \hline
 \multicolumn{1}{|c|}{$y_c$}
 & \multicolumn{1}{c|}{$d\sigma/dy$ (pb)}
 & \multicolumn{1}{c|}{$\delta_{\rm stat}$ (\%)}
 & \multicolumn{1}{c||}{$\delta_{\rm sys}$ (\%)}
 & \multicolumn{1}{c|}{$\delta_{\rm unc}$ (\%)}
 & \multicolumn{1}{c|}{$\delta_{\rm 1}$ (\%)}
 & \multicolumn{1}{c|}{$\delta_{\rm 2}$ (\%)}
 & \multicolumn{1}{c|}{$\delta_{\rm 3}$ (\%)}
\\ \hline
 0.05& $6.95\cdot10^{1}$ & $\pm   7.5$        & $\pm 4.1$          & $\pm 3.7$          & $^{- 0.8}_{+ 0.9}$ & $^{- 0.5}_{+ 0.1}$ & $^{- 1.3}_{+ 1.3}$\\
 0.15& $5.92\cdot10^{1}$ & $\pm   6.5$        & $\pm 2.4$          & $\pm 1.2$          & $^{+ 0.3}_{- 0.5}$ & $^{- 0.8}_{+ 0.8}$ & $^{- 1.8}_{+ 1.8}$\\
 0.27& $4.27\cdot10^{1}$ & $\pm   6.5$        & $^{+ 2.5}_{- 2.2}$ & $^{+ 1.9}_{- 1.8}$ & $^{+ 1.2}_{- 0.8}$ & $^{- 0.5}_{+ 0.9}$ & $^{- 0.9}_{+ 0.9}$\\
 0.41& $3.52\cdot10^{1}$ & $\pm   7.6$        & $^{+ 2.2}_{- 2.4}$ & $^{+ 2.1}_{- 2.2}$ & $^{+ 0.5}_{- 0.2}$ & $^{+ 0.3}_{- 0.6}$ & $^{- 0.2}_{+ 0.2}$\\
 0.55& $2.46\cdot10^{1}$ & $\pm   9.6$        & $^{+ 3.6}_{- 3.2}$ & $^{+ 2.8}_{- 2.6}$ & $^{+ 0.4}_{+ 0.2}$ & $^{+ 2.0}_{- 1.4}$ & $^{- 1.1}_{+ 1.1}$\\
 0.69& $1.55\cdot10^{1}$ & $^{+ 15.}_{- 13.}$ & $^{+ 4.6}_{- 6.7}$ & $^{+ 3.2}_{- 3.4}$ & $^{- 0.9}_{- 1.0}$ & $^{+ 3.1}_{- 5.6}$ & $^{- 1.0}_{+ 1.0}$\\
 0.83& $1.49\cdot10^{1}$ & $^{+ 18.}_{- 16.}$ & $^{+ 8.9}_{- 8.8}$ & $^{+ 5.6}_{- 5.8}$ & $^{+ 1.7}_{+ 0.5}$ & $^{+ 5.8}_{- 5.7}$ & $^{- 3.4}_{+ 3.4}$\\ \hline
\end{tabular}
\end{center}
\end{table}
%
%
\clearpage
\begin{table}
\rotatebox{90}{
\begin{minipage}[b]{\textheight}
\caption
{
The double differential cross
section $d\sigma/dxdQ^2$
for the reaction $e^{+} p \rightarrow \bar{\nu}_e X$.
The following quantities are given for each bin:
the $Q^2$ and $x$ values at which the cross section is quoted,
$Q^2_c$ and $x_c$;
the measured Born--level cross section;
the statistical uncertainty;
the total systematic uncertainty;
the uncorrelated systematic uncertainty and
those systematic uncertainties with significant (assumed 100\%)
correlations between cross-seciton bins.
The systematic uncertainties considered to be correlated were:
the FCAL energy scale ($\delta_1$);
the BCAL energy scale ($\delta_2$) and
the uncertainty in the parton-shower scheme ($\delta_3$).
}
\label{tab:xq2corr_1}
\vskip 0.5 cm
\newcommand{\lw}[1]{\smash{\lower1.8ex\hbox{#1}}}\begin{center}
\begin{tabular}{|r|r|c|r|r||r|r|r|r|} \hline
 \multicolumn{1}{|c|}{$Q^2_c$(GeV$^2$)}
 & \multicolumn{1}{|c|}{$x_c$}
 & \multicolumn{1}{c|}{$d^2\sigma/dxdQ^2$ (pb/GeV$^2$)}
 & \multicolumn{1}{c|}{$\delta_{\rm stat}$ (\%)}
 & \multicolumn{1}{c||}{$\delta_{\rm sys}$ (\%)}
 & \multicolumn{1}{c|}{$\delta_{\rm unc}$ (\%)}
 & \multicolumn{1}{c|}{$\delta_{\rm 1}$ (\%)}
 & \multicolumn{1}{c|}{$\delta_{\rm 2}$ (\%)}
 & \multicolumn{1}{c|}{$\delta_{\rm 3}$ (\%)}
\\ \hline
  280& 0.015& $7.0\cdot10^{-1}$  & $^{+ 17.}_{- 15.}$ & $^{+ 9.2}_{- 9.9}$ & $\pm 5.5$          & $^{- 1.8}_{+ 1.3}$ & $^{- 3.6}_{+ 1.1}$ & $^{- 7.2}_{+ 7.2}$\\
  280& 0.032& $1.99\cdot10^{-1}$ & $^{+ 20.}_{- 17.}$ & $^{+ 9.6}_{- 9.1}$ & $^{+ 3.5}_{- 3.9}$ & $^{- 1.1}_{+ 2.3}$ & $^{+ 0.6}_{+ 3.0}$ & $^{- 8.1}_{+ 8.1}$\\
  280& 0.068& $6.2\cdot10^{-2}$  & $^{+ 29.}_{- 24.}$ & $\pm 17.$          & $\pm 16.$          & $^{- 1.4}_{+ 3.1}$ & $^{+ 0.3}_{- 0.1}$ & $^{- 5.0}_{+ 5.0}$\\
  530& 0.015& $3.86\cdot10^{-1}$ & $^{+ 19.}_{- 16.}$ & $^{+ 8.6}_{- 10.}$ & $^{+ 7.9}_{- 8.4}$ & $^{- 2.3}_{+ 0.8}$ & $^{- 3.2}_{- 1.0}$ & $^{- 3.4}_{+ 3.4}$\\
  530& 0.032& $1.62\cdot10^{-1}$ & $^{+ 16.}_{- 14.}$ & $\pm 5.4$          & $\pm 5.0$          & $^{- 0.4}_{- 1.0}$ & $^{- 1.6}_{+ 1.9}$ & $^{- 0.6}_{+ 0.6}$\\
  530& 0.068& $6.4\cdot10^{-2}$  & $^{+ 18.}_{- 15.}$ & $^{+ 4.5}_{- 3.7}$ & $^{+ 1.9}_{- 2.2}$ & $^{- 1.9}_{+ 3.2}$ & $^{+ 0.5}_{+ 0.8}$ & $^{- 2.3}_{+ 2.3}$\\
  530& 0.130& $2.41\cdot10^{-2}$ & $^{+ 28.}_{- 22.}$ & $^{+ 5.5}_{- 3.1}$ & $^{+ 3.8}_{- 2.3}$ & $^{- 0.4}_{+ 2.9}$ & $^{+ 1.5}_{+ 0.5}$ & $^{+ 1.9}_{- 1.9}$\\
  950& 0.015& $2.97\cdot10^{-1}$ & $^{+ 19.}_{- 17.}$ & $^{+ 10.}_{- 8.5}$ & $^{+ 8.8}_{- 8.2}$ & $^{+ 1.7}_{+ 3.5}$ & $^{- 1.6}_{+ 2.8}$ & $^{- 1.4}_{+ 1.4}$\\
  950& 0.032& $1.53\cdot10^{-1}$ & $^{+ 12.}_{- 11.}$ & $^{+ 2.8}_{- 4.4}$ & $^{+ 1.9}_{- 2.0}$ & $^{- 0.4}_{+ 0.0}$ & $^{- 3.3}_{+ 0.4}$ & $^{+ 2.0}_{- 2.0}$\\
  950& 0.068& $5.59\cdot10^{-2}$ & $^{+ 14.}_{- 12.}$ & $^{+ 2.2}_{- 2.0}$ & $^{+ 1.6}_{- 1.0}$ & $^{- 0.1}_{- 0.1}$ & $^{- 1.0}_{+ 0.6}$ & $^{- 1.4}_{+ 1.4}$\\
  950& 0.130& $2.61\cdot10^{-2}$ & $^{+ 18.}_{- 15.}$ & $^{+ 3.8}_{- 3.6}$ & $^{+ 1.4}_{- 1.3}$ & $^{+ 0.6}_{+ 0.2}$ & $^{- 0.8}_{+ 1.2}$ & $^{- 3.3}_{+ 3.3}$\\
  950& 0.240& $5.7\cdot10^{-3}$  & $^{+ 35.}_{- 27.}$ & $^{+ 5.4}_{- 4.9}$ & $^{+ 3.2}_{- 3.1}$ & $^{+ 1.3}_{+ 0.0}$ & $^{+ 1.5}_{+ 0.4}$ & $^{- 3.9}_{+ 3.9}$\\
 1700& 0.032& $6.9\cdot10^{-2}$  & $^{+ 15.}_{- 13.}$ & $^{+ 4.3}_{- 3.1}$ & $^{+ 3.2}_{- 2.6}$ & $^{- 0.1}_{+ 0.8}$ & $^{+ 2.1}_{+ 0.9}$ & $^{+ 1.8}_{- 1.8}$\\
 1700& 0.068& $3.03\cdot10^{-2}$ & $^{+ 14.}_{- 13.}$ & $^{+ 1.6}_{- 2.2}$ & $^{+ 0.5}_{- 0.8}$ & $^{+ 0.6}_{+ 0.2}$ & $^{- 1.6}_{+ 0.0}$ & $^{+ 1.4}_{- 1.4}$\\
 1700& 0.130& $1.31\cdot10^{-2}$ & $^{+ 19.}_{- 16.}$ & $^{+ 1.1}_{- 1.9}$ & $^{+ 0.8}_{- 1.5}$ & $^{- 0.2}_{- 0.7}$ & $^{+ 0.8}_{- 1.0}$ & $^{+ 0.2}_{- 0.2}$\\
 1700& 0.240& $5.1\cdot10^{-3}$  & $^{+ 23.}_{- 19.}$ & $^{+ 2.6}_{- 1.9}$
 & $^{+ 1.1}_{- 0.7}$ & $^{+ 2.1}_{- 1.7}$ & $^{+ 1.1}_{- 0.1}$ & $^{+
 0.5}_{- 0.5}$\\ \hline
\end{tabular}
\end{center}
\end{minipage}
}
\end{table}
%
%
\clearpage
\addtocounter{table}{-1}
\begin{table}
\rotatebox{90}{
\begin{minipage}[b]{\textheight}
\caption
{
continued.
}
\label{tab:xq2corr_2}
\vskip 0.5 cm
\newcommand{\lw}[1]{\smash{\lower1.8ex\hbox{#1}}}\begin{center}
\begin{tabular}{|r|r|c|r|r||r|r|r|r|} \hline
 \multicolumn{1}{|c|}{$Q^2_c$(GeV$^2$)}
 & \multicolumn{1}{|c|}{$x_c$}
 & \multicolumn{1}{c|}{$d^2\sigma/dxdQ^2$ (pb/GeV$^2$)}
 & \multicolumn{1}{c|}{$\delta_{\rm stat}$ (\%)}
 & \multicolumn{1}{c||}{$\delta_{\rm sys}$ (\%)}
 & \multicolumn{1}{c|}{$\delta_{\rm unc}$ (\%)}
 & \multicolumn{1}{c|}{$\delta_{\rm 1}$ (\%)}
 & \multicolumn{1}{c|}{$\delta_{\rm 2}$ (\%)}
 & \multicolumn{1}{c|}{$\delta_{\rm 3}$ (\%)}
\\ \hline
 3000& 0.068& $2.12\cdot10^{-2}$ & $^{+ 13.}_{- 12.}$ & $^{+ 1.1}_{- 5.1}$ & $^{+ 0.6}_{- 1.9}$ & $^{- 1.3}_{- 1.9}$ & $^{+ 0.5}_{- 4.1}$ & $^{- 0.8}_{+ 0.8}$\\
 3000& 0.130& $8.0\cdot10^{-3}$  & $^{+ 18.}_{- 15.}$ & $^{+ 1.9}_{- 1.6}$ & $^{+ 0.6}_{- 0.5}$ & $^{+ 1.1}_{- 1.3}$ & $^{+ 0.7}_{+ 1.0}$ & $^{- 0.8}_{+ 0.8}$\\
 3000& 0.240& $2.79\cdot10^{-3}$ & $^{+ 23.}_{- 19.}$ & $^{+ 2.8}_{- 3.2}$ & $^{+ 0.9}_{- 0.8}$ & $^{+ 2.5}_{- 2.9}$ & $^{- 0.8}_{+ 0.6}$ & $^{+ 0.3}_{- 0.3}$\\
 3000& 0.420& $9.5\cdot10^{-4}$  & $^{+ 33.}_{- 26.}$ & $^{+ 4.5}_{- 6.3}$ & $^{+ 0.9}_{- 1.4}$ & $^{+ 3.9}_{- 5.7}$ & $^{+ 0.4}_{- 0.9}$ & $^{- 2.1}_{+ 2.1}$\\
 5300& 0.068& $6.3\cdot10^{-3}$  & $^{+ 27.}_{- 22.}$ & $^{+ 5.5}_{- 5.8}$ & $^{+ 1.6}_{- 1.3}$ & $^{+ 1.5}_{+ 2.2}$ & $^{+ 4.3}_{- 5.4}$ & $^{- 1.5}_{+ 1.5}$\\
 5300& 0.130& $3.75\cdot10^{-3}$ & $^{+ 20.}_{- 17.}$ & $^{+ 6.1}_{- 5.1}$ & $^{+ 1.3}_{- 0.4}$ & $^{+ 1.3}_{- 0.8}$ & $^{+ 5.2}_{- 4.4}$ & $^{+ 2.5}_{- 2.5}$\\
 5300& 0.240& $1.51\cdot10^{-3}$ & $^{+ 24.}_{- 20.}$ & $^{+ 4.9}_{- 5.8}$ & $^{+ 0.6}_{- 0.7}$ & $^{+ 4.8}_{- 5.7}$ & $^{+ 0.8}_{- 1.1}$ & $^{- 0.3}_{+ 0.3}$\\
 5300& 0.420& $2.7\cdot10^{-4}$  & $^{+ 49.}_{- 35.}$ & $^{+ 9.8}_{- 7.6}$ & $^{+ 1.3}_{- 0.5}$ & $^{+ 9.6}_{- 7.6}$ & $^{+ 0.6}_{+ 0.9}$ & $^{+ 0.0}_{+ 0.0}$\\
 9500& 0.130& $1.06\cdot10^{-3}$ & $^{+ 35.}_{- 27.}$ & $\pm 12.$          & $^{+ 1.4}_{- 1.5}$ & $^{+ 0.7}_{- 1.3}$ & $^{+ 12.}_{- 11.}$ & $^{+ 3.8}_{- 3.8}$\\
 9500& 0.240& $5.6\cdot10^{-4}$  & $^{+ 30.}_{- 23.}$ & $^{+ 6.4}_{- 7.7}$ & $\pm 0.7$          & $^{+ 4.9}_{- 5.4}$ & $^{+ 3.7}_{- 5.2}$ & $^{+ 1.6}_{- 1.6}$\\
 9500& 0.420& $1.82\cdot10^{-4}$ & $^{+ 43.}_{- 31.}$ & $\pm 11.$          & $^{+ 0.5}_{- 0.6}$ & $\pm 11.$          & $^{- 0.3}_{- 0.2}$ & $^{- 0.5}_{+ 0.5}$\\
17000& 0.240& $8.0\cdot10^{-5}$  & $^{+ 69.}_{- 44.}$ & $^{+ 21.}_{- 19.}$ & $^{+ 1.8}_{- 1.7}$ & $^{+ 1.0}_{- 1.2}$ & $^{+ 20.}_{- 19.}$ & $^{+ 3.5}_{- 3.5}$\\
17000& 0.420& $6.3\cdot10^{-5}$  & $^{+ 54.}_{- 37.}$ & $^{+ 16.}_{- 14.}$ & $\pm 0.9$          & $^{+ 14.}_{- 13.}$ & $^{+ 4.6}_{- 3.3}$ & $^{+ 3.5}_{- 3.5}$\\ \hline
\end{tabular}
\end{center}
\end{minipage}
}
\end{table}
%
%
\clearpage
%
%
%
%
%
%
\newpage
%
%
%
%
%
%
%
%
\end{document}